\documentclass[12pt]{article}
\usepackage[english]{babel}
\usepackage{amssymb}
\usepackage{epsfig,color}
\usepackage[table]{xcolor}
\usepackage{pstricks,graphicx,epsfig,color,amssymb,amsmath,amscd}
\usepackage{cite}
\usepackage{graphicx}
\usepackage{makeidx}
\usepackage{amsmath}
\usepackage{nccmath}
\usepackage{array}
\usepackage{setspace}
\usepackage{float}
\usepackage{subcaption}
\graphicspath{ {./pictures/}} 
\usepackage{capt-of}
\usepackage{lipsum}
\usepackage{verbatim}
\usepackage[]{caption}
\def\mn{{\mu\nu}}

\renewcommand\rho{\varrho}

\newcommand{\be}{\begin{eqnarray}}
	\newcommand{\ee}{\end{eqnarray}}
\newcommand{\rar}{\rightarrow}

\newcommand{\tcviol}{\textcolor{violet}}

\numberwithin{equation}{subsection}
\providecommand{\keywords}[1]
{
  \small	
  \textbf{\textit{Keywords:  \\}} #1
}

\begin{document}
\selectlanguage{english}
	\begin{titlepage}
		\title{Graviton to photon conversion in curved space-time and external magnetic field}
		\author{A. D. Dolgov$^{a, b}$, L. A. Panasenko$^{a}$, V. A. Bochko$^{a}$}
				
		\maketitle
		\begin{center}
			$^a${Department of Physics, Novosibirsk State University, \\Pirogova 2, Novosibirsk 630090, Russia}\\
			$^b${{Bogoliubov Laboratory of Theoretical Physics, JINR, Dubna, 141980, Russia}}
			Correspondence: dolgov@g.nsu.ru (A. D.), l.vetoshkina@g.nsu.ru (L. A.), v.bochko@g.nsu.ru (V. A.)
		\end{center}
		
		\begin{abstract}
			{The suppression of relic gravitational waves due to their conversion into electromagnetic radiation in a 
			cosmological magnetic field is studied. The coupled system of equations describing 
			gravitational and electromagnetic wave propagation in an arbitrary curved space-time and in external magnetic field
			is derived.  The subsequent elimination of photons from the beam due to their interaction with the primary plasma is taken into account.
			The resulting system of equations is solved numerically in Friedman-LeMaitre-Robertson-Walker metric for 
			the upper limit of the intergalactic magnetic field strength of 1 nGs. We conclude that the gravitational wave conversion 
			into photons in the intergalactic magnetic field cannot significantly change the amplitude of the relic gravitational wave and
			their frequency spectrum.} 			
		\end{abstract}
		
		\thispagestyle{empty}		
	\end{titlepage}
	
	\tableofcontents
	\newpage
	\section{Introduction}
	The transformation between gravitons and photons in external magnetic field was considered in multitude of 
	papers starting from 1961~\cite{g-to-gam1} -\cite{g-to-gam7}. The problem acquired particular importance in
	connection with {the} possible transformation of relic gravitational waves  (GW) produced at 
	 {the} inflationary stage into electromagnetic waves (EMW) in primordial magnetic fields.
	However, in all previous works see e.g.~\cite{g-to-gam7,g-to-gam8} the calculations have been always 
	done in Minkowski space-time, though the curvature effects in the very early universe could be quite 
	essential.
	 
	{In the present work we go beyond the flat space-time restriction and consider graviton and photon 
	propagation in arbitrary curved backgroud.
	The propagation of gravitational waves in curved space-time was almost always considered in {Friedman-LeMaitre-Robertson-Walker metric (FLRW)},
	see e.g. textbooks~\cite{mtw,GR-2}, except for some Bianchi types metrics and
	our recent paper~\cite{OurWork}, where an arbitrary background metric was allowed. 	
	Here we derive the propagatoin equations for the coupled system of
	photons and gravitons in  {an} arbitrary background. Next we will turn to the
	Friedman--LeMaitre-Robertson-Walker (FLRW) space-time, {that is a good approximation to the real universe. 
	However, deviations from FLRW could be essential and lead
	to interesting observable effects.}}
	
	{Already for a century the Friedman equations serve as a basement for the conventional cosmological model.
	They perfectly well describe the early universe, that is homogeneous and isotropic to a very good approximation.
	They are operative also in the contemporary universe on very large scales.
	Friedman cosmology allows for description of cosmological dark matter and what's more surprising dark energy,
	though the physical nature of the latter is not yet established.
	}

	{The propagation of gravitational waves in curved space-time was almost always considered in FLRW metric, except for some Bianchi types metrics and
	our recent paper~\cite{OurWork}, where an arbitrary background metric was allowed. }	
		
	Here we derive {the coupled}
	equations of motion for  metric perturbations and  electromagnetic waves over {an} arbitrary cosmological 
	background in {the} external cosmic magnetic field. {The} metric perturbations and EMW
	are treated in the first order of {the} perturbation theory. 
	
	{We introduce the full electromagnetic
field $\bar A_{\mu}$ as the sum of an external classical
component of the electromagnetic field $A_{\mu}$ and a small quantum
fluctuation $f_{\mu}$, which is considered as a perturbation,
\begin{align}
 \bar A_{\mu} =  A_{\mu} +  f_{\mu}.
 \end{align}
Then, the stress tensors of $\bar A_{\mu},  A_{\mu}$ and $f_{\mu}$ are introduced accordingly:
\be
	\bar F_{\mu,\nu} &=& \partial_\mu \bar A_\nu - \partial_\mu \bar A_\nu\\
    	F_{\mu\nu} &=& \partial_\mu  A_\nu - \partial_\nu A_\mu,\\
	f_{\mu\nu}&=&\partial_\mu f_\nu - \partial_\nu f_\mu.
\label{def-F-f}
\ee
}
{
The full metric tensor $\bar g_{\mu\nu}$ is expanded around
the metric tensor of the background space-time  $g_{\mu\nu}$
as 
\be
\bar g_{\mu\nu} = g_{\mu\nu} + h_{\mu\nu},\nonumber\\
\bar g^{\mu\nu} = g^{\mu\nu} -  h^{\mu\nu}
\label{bar-g}
\ee 
with $h_{\mu\nu}$ being a small perturbation of the metric.
}

{The properties of the metric tensor $g_{\mu\nu}$ are specified by:
 the orthogonality condition $g_{\mu\nu} g^{\mu\lambda} = \delta_\mu^\lambda$, where $\delta_\mu^\lambda $ is the Kronecker delta-symbol;
rising and lowering of the indices of the tensors $h_{\mu\nu}$ and $f_{\mu\nu}$ by the background metric tensor $g^{\mu\nu}$.
Note, that the indices of the full and classical stress tensors of
the electromagnetic fields are raised and lowered with the full metric tensor $\bar g_{\mu\nu}$.
}
		
	The corrections to the metric determinant $\bar g$ can be found from the first order expansion of an
	arbitrary non-degenerate matrix $\cal M$:
	\be
	\det[ {\cal M}  +\delta {\cal M} ] = \det[ {\cal M}] \left(1 + {Tr  [{\cal M}^{-1} \delta {\cal M} ]}\right).
	\label{det-g}
	\ee
	So we obtain: 
	\be 
	\det [\bar g_{\mu\nu} ] = \det [ g_{\mu\nu} h_{\mu\nu}] = \det [g_{\mu\nu} ]  \left( 1 + g^{\mu\nu} h_{\mu\nu} \right) . 
	\label{det-g-expan}
	\ee
	It is assumed usually that tensor perturbations are traceless:
	\be
	h \equiv g^{\mu\nu} h_{\mu\nu} = 0.
	\label{h1}
	\ee
	However, we see in what follows that {the} corrections to the Maxwell energy-momentum tensor are not
	traceless, see e.g. {eqs.~(\ref{trace-T-HE1}, \ref{T-mu-mu-anom})} 
	and a nonzero trace of the gravitational field source leads to {the} nonzero $h$, so
	$\det [\bar g_{\mu\nu}]  = \det [g_{\mu\nu} ]  \left( 1 +  h \right) $.

	The initially derived equations  are supposed to be valid in an arbitrary space-time metric, but
	ultimately we assume that the background metric has the $3D$-flat FLRW form:
	\be
	ds^2 = dt^2 - a^2(t) \sum_{j=1,2,3}  dx^2_j
	\label{ds-2}
	\ee
	where $a(t)$ is the cosmological scale factor. The Hubble parameter is expressed through it in the usual way
	as $H =\dot a/a$.   {The curved metric reduces to the flat {one when} $a \to 1$}.

The paper is organized as follows. We start in {Sec.\ref{ArbST} from a brief reminding} of 
{the} equation for metric perturbations over  arbitrary space-time. In Sec.\ref{Helis} we {recall}
the expansion of  metric perturbations in {terms of helicity eigenstates.} 
After that in Sec.\ref{Mixing} we show that {the} scalar and tensor modes can mix  in general case of inhomogeneous space. 
{Further, in Sec.\ref{GWprop}  we consider propagation of metric perturbation over FLRW space-time in external magnetic field. 
In Sec.\ref{EMWprop} the propagation of electromagnetic waves in magnetic field is considered.}
In both sections we start from the classical Maxwell and Hilbert-Einstein actions ignoring for a while the 
Heisenberg-Euler (HE)~\cite{HE} corrections, {the} quantum trace anomaly and matter effects.
{They are taken into} account step by step in the subsequent subsections. 
{On the way we discuss the definition of physical magnetic fields through the electromagnetic field tensor $F_{\mu\nu}$ in cosmological 
background (subsection~\ref{ss-maxwell-B}) and the impact of the HE-corrections to the electromagnetic wave propagation expressed through the physical magnetic field $ \bf{B}$ 
(subsection~ \ref{ss-eq-of-mot-HE}). } In Sec.\ref{WholeSDE} we analyze the {full set}   of 
differential equations (SDE) for {$(g-\gamma)$-coupled} system, choose a reference {frame}, 
and simplify the system for the {choice made}. Next, in Sec.\ref{TwoCases}, we divide the task into two cases: $\bf{k}||\bf{B}$ and 
$\bf{k}\bot\bf{B}$, and find out that  the conversion effect is present only for the second case. In the last Sec.\ref{Sol} we divide 
SDE into two independent subsystems and solve the first of them numerically. In conclusion we summarize the obtained results and
 formulate the prospects for the {future} research.

	\section{Metric perturbations in general case}
	\label{ArbST}
	In \cite{OurWork} we obtained equation (23) for the propagation of metric perturbations in arbitrary space-time. Let's write it for two lower indices:
	\be \label{EOM}
	 &D^2 h_{\mu\nu}-2h^{\alpha\beta}R_{\alpha\mu\nu\beta}-\left(h_{\alpha\mu}R^{\alpha}_{\nu}+h_{\alpha\nu}R^{\alpha}_{\mu}\right)+h_{\mu\nu}R-g_{\mu\nu}\left(h^{\alpha\beta}R_{\alpha\beta}+\frac{1}{2}D^2 h\right)\nonumber\\
	&=-2 (8\pi G)T^{(1)}_{\mu\nu},
	\ee
	{where $G$ is the gravitational constant, $D^2=D^{\alpha}D_{\alpha}$ and $D_{\alpha}$ is the covariant derivative, $R_{\alpha\mu\nu\beta}, R_{\mu\nu}, R$ are the Riemann tensor, Ricci tensor and scalar curvature respectively.}

	The equation contains additional terms that disappear in 
	the special cases of Minkowski and FLRW spaces. 
	These extra terms could have significant effects on the GW and EMW propagation over
	 background metric that differs from the FLRW one.

	Let us note the agreement between equation (\ref{EOM}) and equation (2.33) from the work  \cite{Gasperini}, 
	published {after} our work \cite{OurWork}. The apparent difference with our result disappears in the Lorentz calibration
	\be\label{LorGauge}
	D_{\mu}h^{\mu}_{\nu}=\frac{1}{2}\partial_{\nu}h.
	\ee
	In this article, the authors obtained the same equation using a double variation of the action, while we obtained it by expanding the Einstein equation to first order in perturbation.

	\section{Helicity decomposition and choice of gauge}
	\label{Helis}
Now it is worth recalling the formalism of the expansion of the perturbation $h_{\mu\nu}$ in {terms of helicity states.}
The generally accepted approach is that (along with the vectors $C_i, G_i$ and the traceless tensor $D_{ij}$) four scalars $A, B, E, F$ are introduced, through which the components of the {metric perturbation}  are expressed:
	\be
	\label{Decom2}
	&&h_{00}=-E,\\
	&&h_{0i}=a\left(\frac{\partial F}{\partial x^i}+G_i\right),\\
	&&h_{ij}=a^2\left(A\delta_{ij}+\frac{\partial^2 B}{\partial x^i \partial x^j}+\frac{\partial C_i}{\partial x^j}+\frac{\partial C_j}{\partial x^i}+D_{ij}\right).
	\label{Decom3}
	\ee
{One can impose gauge conditions such that two scalars turn to zero. The so called synchronous gauge 
corresponds to the choice  $E=0$ and $F=0$. Under this gauge there still remains some more 
freedom, that may allow to simplify algebra in a specific problem.} The second well-known type of gauge is 
{the} Newtonian gauge, where  {$B=0$, $F=0$}, $E\equiv 2\Phi$, and $A\equiv-2\Psi$. 
The choice of this gauge {better fits}
our task, so for the scalar sector we will use the Newtonian gauge.
	
	In addition to the gauge in the scalar sector, the Lorentz gauge (\ref{LorGauge}) is usually imposed on the entire 
	{tensor perturbation  of metric. This calibration naturally arises in the case when the so-called harmonic Fock 
	coordinates are used.
	It allows  to obtain a simpler expression for the Ricci tensor and, as a consequence, to simplify the equation for the propagation of 
	metric   perturbations. 
	Formerly in our paper \cite{OurWork}, we have only used the Lorentz gauge.}
	
	\section{Mixing of metric perturbation modes \label{Mixing} }
	Note that from the expression for the trace of the equation (\ref{EOM}) it turns out that in the general case, for an 
	arbitrary form of the Ricci tensor, {there appears a} 
	mixing of scalar and tensor modes of metric perturbations. 
	{In general} case it is impossible to separate the equations for these two sectors. Indeed, 
	{taking trace of eq. (\ref{EOM}),} we obtain the following expression
	\be\label{EOMtrace}
	\partial^2h+4h^{\alpha\beta}R_{\alpha\beta}-hR=16\pi G T^{\alpha (1)}_{\alpha},
	\ee
	{where $\partial^2=\partial^{\mu}\partial_{\mu}$}, and from which it is clearly seen that {the second term in the left hand side includes
	both terms from the scalar and the tensor sectors.}
	
	In addition, it is important to pay attention to the trace from the source {in the right hand} side of the equation. 
	As it will be shown below for the problem of the graviton conversion  into photons in an external magnetic field, 
	the trace from the correction to the EMT contains a convolution of the background electromagnetic tensor and the 
	{tensor perturbation to the  metric,} $h_{\mu\nu}$, which also leads to mixing {between}  scalar and tensor modes.
	This result is {evident}, 
	because the expansion of metric perturbations in polarizations is valid for a problem with axial
	 symmetry: in this problem there is only one  {specific} direction - the direction of the wave vector $\bf{k}$ 
	 of the metric perturbation.
	 If space is for some reason {unisotropic} (as, for example, in the case of an external magnetic field or in the 
	 presence of an anisotropic stress tensor), this symmetry disappears.
	
	\section{Metric perturbations in magnetic field}
	\label{GWprop} 
	\subsection{Equation in the FLRW metric}
	Recall that in the FLRW metric
	\be\label{Ricci}
	&R_{00}=-3\frac{\ddot{a}}{a},\\
	&R_{ij}=-g_{ij}\left(\frac{\ddot{a}}{a}+2H^2\right).\\
	&R=-6\left(\frac{\ddot{a}}{a}+H^2\right).
	\ee
	We write the trace of the GW tensor in the following form:
	\be\label{hii}
	h=g^{\mu\nu}h_{\mu\nu}=h_{00}-\frac{h_{xx}+h_{yy}+h_{zz}}{a^2}\equiv h_{00}+h^i_i,
	\ee
	where the {notation  $h^i_i=-\left(h_{xx}+h_{yy}+h_{zz}\right)/{a^2}$ was introduced:}
	
	Let us write down the system of equations (\ref{EOM}) for the case of the FLRW metric. To do this, we will use the expressions (\ref{Ricci}-\ref{hii}). We get
	\be\label{EOM2}
	&D^2h_{\mu\nu}-2h^{\alpha\beta}R_{\alpha\mu\nu\beta}-g_{\mu\nu}\left[\frac{1}{2}\partial^2h-3\frac{\ddot{a}}{a}h_{00}-\left(\frac{\ddot{a}}{a}+2H^2\right)h^i_i\right]-6H^2h_{00}\delta_{0\mu}\delta_{0\nu}\nonumber\\
	&{-2\left(\frac{2\ddot{a}}{a}+H^2\right)h_{\mu\nu}\left[1-\delta_0^{\mu}\delta_0^{\nu}\right]}=-16\pi G T^{(1)}_{\mu\nu},
	\ee
	{where the Latin indices are the spatial ones (vary from one to three).}
	
	For a medium {where the} perturbation propagates, we will consider a model of an ideal fluid. 
	The total energy-momentum tensor in this case is determined through the full metric as follows
	\be\label{IdLicEMT}
	\overline{T}^{medium}_{\mu\nu}=-p \overline{g}_{\mu\nu} +(p+\rho)\overline{u}_{\mu}\overline{u}_{\nu},
	\ee
	where $\rho$ is the energy density, $p$ is the pressure, $u_{\mu}$ is the four-speed. Then the right side of 
	equation (\ref{EOM2}) can be rewritten as
	\be
	-16\pi G T^{(1)}_{\mu\nu} 
	= -16\pi G \left(T^{medium(1)}_{\mu\nu}+T^{EM(1)}_{\mu\nu}\right)=16\pi G p h_{\mu\nu} - 16\pi G T^{EM(1)}_{\mu\nu},
	\ee
	where the first term in the last equality is obtained by expanding equation (\ref{IdLicEMT}) to the first order in perturbation at {$u_{j}=0$ (index j varies from one to three)} and the second term is responsible for the perturbation of the EMT due to the presence of an external magnetic field.
	
	The factor before the last term on the left side of the equation (\ref{EOM2}) is exactly
	\be
	-2\left(2\frac{\ddot{a}}{a}+H^2\right)=16\pi G p.
	\ee

	Thus, equation (\ref{EOM2}) can be simplified:
	\be\label{EOM3}
	&D^2h_{\mu\nu}-2h^{\alpha\beta}R_{\alpha\mu\nu\beta}-g_{\mu\nu}\left[\frac{1}{2}\partial^2h-3\frac{\ddot{a}}{a}h_{00}-\left(\frac{\ddot{a}}{a}+2H^2\right)h^i_i\right]-6H^2h_{00}\delta_{0\mu}\delta_{0\nu}\nonumber\\
	&=-16\pi G T^{EM(1)}_{\mu\nu},
	\ee
	
	{Now, for brevity, we omit the expressions for the components of the 
	Riemann tensor and the Christoffel symbols in the 
	covariant derivative and write the final equations for $00$, $0j$, and $ij$ components separately:}
	\be\label{GWsys1}
	&\left[\partial_t^2+3H\partial_t-\frac{\Delta}{a^2}+3\left(\frac{\ddot{a}}{a}-4H^2\right)\right]h_{00}-\frac{1}{2}\partial^2h+\left(4H^2-\frac{\ddot{a}}{a}\right)h^i_i=-16\pi G T^{EM(1)}_{00},\nonumber\\
	&2H\left[\partial_j h_{00}+\frac{\partial_x h_{xj}+\partial_y h_{yj}+\partial_z h_{zj}}{a^2}\right]=-16\pi G T^{EM(1)}_{0j},\nonumber\\
	&\left[\partial_t^2+3H\partial_t-\frac{\Delta}{a^2}\right]h^i_{j}+\delta^i_{j}\left[-\frac{\partial^2 h}{2}+\left(\frac{\ddot{a}}{a}+2H^2\right)h_{00}+\frac{\ddot{a}}{a}h^l_l\right]=-16\pi G T^{i\,EM(1)}_{j},\,\,\,\,\,
	\ee
	where {notation} (\ref{hii}) is used, and the last equation is written in {terms of}
	mixed components, since then it looks more consistent with the equation for the $00$ component.

	\subsection{Corrections to the energy-momentum tensor (EMT)}
	Corrections to the EMT  are due to the presence of an external electromagnetic field. We will find them in accordance with the definition of EMT of matter:
	\be
	\overline{T}_{\mu\nu}=\frac{2}{\sqrt{-\overline{g}}}\frac{\delta \overline{\cal{A}}_{mttr}}{\delta \overline{g}_{\mu\nu}}.
	\ee
	There are two contributions to the EMT perturbation: from the Maxwell action and from the Heisenberg-Euler action. 
	
	{The gravity of the background magnetic field is negligible compared to the background of matter and we ignore its contribution to the EMT corrections}.
	
	\subsubsection{Corrections to EMT emerging from the Maxwell action}
	The Maxwell action is written as follows:
	\be
	{\cal A}{_{Max}} = -\frac{1}{4}\int d^4x  \sqrt{- \bar g} \left( \bar F^2 + \bar A_\mu \bar J^\mu \right),
	\label{A-Max}
	\ee
	where $\bar F^2 = \bar F_{\mu\nu} \bar F^{\mu\nu} = 
	\bar g^{\mu\alpha} \bar g^{\nu\beta} \bar F_{\mu\nu} \bar F_{\alpha\beta} $.	
	Hence the energy-momentum tensor is
	\be
	\bar T_{\mu\nu}^{(Max)} =   \frac{1}{4} \,\bar g_{\mu\nu} \bar F^2 - 
	\bar F_{\mu\alpha} \bar F_{\nu\beta} \, \bar g^{\alpha\beta} .
	\label{T-Max-tot}
	\ee
	or for the mixed components:
	\be
	\bar T_{\nu}^{\mu (Max)} =   \frac{1}{4} \,\delta^{\mu}_{\nu} \bar F^2 -
	\bar F^{\mu}_{.\, \alpha} \bar F_{\nu}^{.\,\alpha} .
	\label{T-Max-mixed}
	\ee
	Clearly this tensor is conserved and its  trace is zero:
	\be
	\bar g^{\mu\nu} \bar T_{\mu\nu}^{(Max)} 
	\equiv \delta^{\mu}_{\nu} \bar T_{\mu}^{\nu(Max)} = 0 .
	\label{trace-T-Max}
	\ee
	The vanishing of the EMT trace in Maxwell electrodynamics is a consequence of
	the conformal invariance of the Maxwell action (\ref{A-Max}). This is not so for higher order quantum
	corrections (trace anomaly), see subsecton~\ref{ss-conf-anom}.

	The trace of the zero order term with mixed components:
	\be
	T_{\nu}^{\mu (Max\, 0)} =   \frac{1}{4} \,\delta^{\mu}_{\nu}  F^2 - 
	F^{\mu}_{.\, \alpha}  F_{\nu}^{.\,\alpha} 
	\label{T-Max-mixed}
	\ee
	is also zero. Note that moving indices up or down in this equation is done by the background 
	metric, e.g.  $F^{\mu}_{.\, \alpha}  F_{\nu}^{.\,\alpha} = 
	g^{\mu\sigma} g^{\alpha\lambda} F_{\sigma\alpha} F_{\nu\lambda}$
	and $F^2 = F_{\alpha\beta}F_{\sigma\lambda} \,g^{\alpha \sigma} g^{\beta \lambda}$. 
	
	The zero order term is presumably small in comparison with the total cosmological energy-momentum
	tensor and can be neglected in what follows.
	
	For the first order term with mixed components we obtain:
	\be
	T_{\nu}^{\mu (Max\,1)} = \frac{1}{2} \delta_{\nu}^{\mu}\left[(F f) -  (FFh) \right] 
	+ h^{\mu\sigma}  F_{\sigma\alpha} F_{\nu}^{.\,\alpha} +
	h^{\alpha\lambda} F^{\mu}_{.\,\alpha} F_{\nu\lambda}
	- f^{\mu}_{.\, \alpha}  F_{\nu}^{.\,\alpha} -F^{\mu}_{.\, \alpha}  f_{\nu}^{.\,\alpha} ,
	\label{T-Max-1}
	\ee
	where $(Ff) = F_{\alpha\beta} f^{\alpha\beta}$ and 
	$(FFh) = F_{\alpha\beta}F_\sigma^{.\,\beta} h^{\alpha\sigma}$.
	Evidently $T_{\mu}^{\mu (Max\,1)} =0$, as is expected.

	\subsubsection{Heisenberg-Euler (HE) Lagrangian}
	The second origin of EMT corrections is Heisenberg-Euler effective Lagrangian~\cite{HE}. It
	describes quartic self-interaction of electromagnetic field and is induced  
	by the loop of virtual electrons with four external electromagnetic legs. In the weak field limit,
	and low energies, much smaller than the electron mass, $m_e$, the corresponding action has the form:
	\be
	{\cal A}_{HE}^{(0)} =  \int d^4x \sqrt{-g} \, C_0\left[(F_{\mu\nu} F^{\mu\nu})^2+
	\frac{7}{4}({\tilde F}^{\mu\nu} F_{\mu\nu})^2 \right] .
	\label{A-HE}
	\ee
	Here  
	\be\label{C0definition}
	C_0 = \alpha^2/(90 m_e^4)
	\ee
	and $\alpha = 1/137$ is the fine structure constant.
	At high temperatures $C$, $\alpha$, and $m_e$ change with $T$.
	At this stage we omit the bar over $F_{\mu\nu}$ to simplify notations. The dual Maxwell tensor is defined as
	\be 
	\tilde F_{\alpha\beta} = \frac{\sqrt{-g}}{2}\,  \epsilon_{\alpha\beta\mu\nu} F^{\mu\nu}, \,\,\,
	\tilde F^{\alpha\beta} = \frac{1}{2 \sqrt{-g}} \epsilon^{\alpha\beta\mu\nu} F_{\mu\nu}, 
	\label{tilde-F-c}
	\ee
	because the tensor quantity is $\sqrt{-g}\,  \epsilon_{\alpha\beta\mu\nu} $ but not just
	$\epsilon_{\alpha\beta\mu\nu}$, see  e.g. chapter 83 from  textbook~\cite{LL-2}.
	
	In what follows we apply this action to photon propagation in external magnetic field $B$ and the weak field limit means 
	$B\ll m_e^2$. 
	
	We need to generalize the Heisenberg-Euler action (\ref{A-HE})
	to  high energies/temperatures and curved  FLRW space-time. To do that let us start
	from the canonical action of photons and electrons  written in {terms of the}
	conformal metric:
	\be 
	ds^2 =  {g_{\mu\nu}^{(c)} dx^\mu dx^\nu=}a^2 (\tau) \left( d\tau^2 - d {\bf r^2} \right) 
	\equiv a^2(\tau) \eta_{\mu \nu} dx^\mu dx^\nu , 
	\label{ds-conf}
	\ee
	where $\eta_{\mu\nu}  = diag [1,-1,-1,-1]$ is the Minkowski metric tensor,  {and $a(\tau)$ is the scale factor 
	{as a function of} conformal time $\tau(t) = \int dt/a(t)$.}
	
	{The action of photons and electrons  written in terms of the inverse metric to $g_{\mu\nu}^{(c)}$,
	given by eq. (\ref{ds-conf}), takes the form: }
	\be{
	{\cal A}_{e\gamma} = \int d^4 x \, a^4 \left[ - \frac{ g^{\mu\alpha}_{c}  g^{\nu\beta}_{c}}{4} F_{\alpha\beta} F_{\mu\nu}+
	\bar \psi \left( i g^{\mu\nu}_{c} \Gamma_\mu \nabla_\nu - m_\psi \right) \psi
	+ e g^{\mu\nu}_{c} A_\mu \bar \psi \Gamma_\nu \psi  \right],}
	\label{A-e-gamma}
	\ee
	where $\Gamma^\mu$ is a generalization of the Dirac $\gamma$ matrices for curved space-time. For FLRW metric
	they have the form $\Gamma_\mu = a \gamma_\mu$, where $\gamma_\mu$ are the
	usual Dirac matrices which {anticommute} as 
	$[\gamma_\mu,\gamma_\nu]= \eta_{\mu\nu}$, $[\Gamma_\mu,\Gamma_\nu]= g_{\mu\nu}$;
	$\nabla_\mu$ is the covariant derivative for spin-(1/2) field. For FLRW metric it has the form 
	$\nabla_\mu = \partial_\mu + (3/2)\partial_\mu (\ln a ) $. 
	
	Introducing conformally transformed spinor $\chi = \psi /a^{3/2}$, we arrive to the action:
	\be
	{\cal A}_{e\gamma} = \int d^4 x  \left[ - \frac{ \eta^{\mu\alpha}  \eta^{\nu\beta}}{4} F_{\alpha\beta} F_{\mu\nu}+
	\bar \chi \left( i  \gamma^\mu \partial_\nu - a m_\psi \right) \chi
	+ a e \eta^{\mu\nu} A_\mu \bar \chi \gamma_\mu \chi  \right].
	\label{A-e-gamma}
	\ee
	This is essentially the same action as it is in flat space-time with rescaled mass and charge: $m\rar a m$ and 
	$e \rar a e$, so formally $C_0 \sim e^4/m^4$ does not change,
	but since we plan to go to very high temperatures, even above the electroweak
	phase transition, when all bare masses of charged particles vanish, we have to substitute the high
	temperature value of the mass, to sum over all charged particles, and to take the high temperature value {of the electromagnetic coupling $\alpha$.} So
	\be 
	C  (T) = \sum_j \frac{\alpha^2 (T) q_j^4}{90 m_j (T)^4} ,
	\label{C-of-T}
	\ee
	where $q_j$ is the charge of the contributing to the loop particles in the electron charge units, e.g.
	for {down or up quarks} $q= -1/3$ or $2/3$.
	
	The integrand in the expression for the action $\cal{A}_{HE}$ is a scalar with respect to the general coordinate 
	transformation, so we can use for it the same expression as (\ref{A-HE}) in arbitrary metric.
	
	{In the early universe at high temperatures the Heisenberg-Euler action keeps the same form
		as (\ref{A-HE}) with substitution of $C(T)$ instead of $C_0$:
	}
	\be
	{\cal A}_{HE} = 
	\int d^4x \sqrt{-\bar g} \, C (T) \left[(\bar F^2)^2+
	\frac{7}{4} (\tilde {\bar F} \bar F)^2 \right],
	\label{A-HE-curv}
	\ee
	where $F^2 = F_{\mu\nu} F^{\mu\nu}$, $\tilde F F = \tilde F_{\mu\nu} F^{\mu\nu}$, and
	we have returned bar over $F_{\mu\nu}$ and to  the metric determinant
	in accordance with expansion (\ref{bar-g}).

	The HE action given by eq.~(\ref{A-HE-curv}) leads to the following contribution to the 
	energy-momentum tensor:
	\be
	T_{\mu\nu}^{HE} = C(T) \left\{ - g_{\mu\nu} \left[( F_{\alpha\beta}  F^{\alpha\beta})^2 -
	\frac{7}{4}({\tilde { F}}^{\alpha\beta}  F_{\alpha\beta})^2 \right] 
	+ 8 ( F_{\alpha\beta}  F^{\alpha\beta})  F_{\mu\lambda} F_\nu^{.\,\lambda}
	\right\}.
	\label{T-HE}
	\ee 
	{Here the over-bars are eliminated to simplify notation but we keep in mind that this expression will be used with
		the non-expanded complete quantities, see  Eq.~(\ref{bar-g}).
	}
	
	An explanatory comment may be in order here, namely,
	the second term containing the dual Maxwell tensor, $\tilde F_{\mu\nu}$, depends upon metric only through the
	factor $(\sqrt{-g})^{-2}$, so with the account of the integration measure
	the action depends on metric as $(\sqrt{-g})^{-1}$ instead
	on $(\sqrt{-g})$. Hence this gives the contribution to $T_{\mu\nu}$ from $(\tilde { F}  F)^2 $
	proportional to $ (+g_{\mu\nu})$
	instead of the usual one proportional to $ (-g_{\mu\nu})$.
	
	One can see that the trace of tensor (\ref{T-HE}) is non-vanishing:
	\be
	T_{\mu}^{\mu\,HE} 
	= C(T) \left[ 4  ( F_{\alpha\beta}  F^{\alpha\beta})^2 +
	{7}({\tilde { F}}^{\alpha\beta}  F_{\alpha\beta})^2 \right] \neq 0 .
	\label{trace-T-HE}
	\ee
	
	It is instructive to check conservation of the energy-momentum tensor (\ref{T-HE}), though it 
	surely must be true, since it was obtained by the variation of a scalar function over metric. 
	Still, at least the verification of the conservation would indicate that the calculations are correct.
	{Let us note that the conservation condition should be fulfilled  only if $C = const$. Evidently
		the energy-momentum tensor (\ref{T-HE}) is non-conserved for a non-constant $C(T)$
		because the dependence on temperature appears due to interaction and an exchange of energy with external system.}
	
	It would be more convenient to express the square of the dual electromagnetic tensor through F. It enters the action in the  
	form, see eq.~(\ref{tilde-F-c}):
	\be
	\left(\tilde{F}F\right)^2 = \tilde{F}^{\mn}F_{\mn} \tilde{F}_{\alpha\beta}F^{\alpha\beta}=\frac{1}{4}\epsilon^{\mu\nu\lambda\sigma}\epsilon_{\alpha\beta\tau\chi}F_{\lambda\sigma}F_{\mn}F^{\tau\chi}F^{\alpha\beta}.
	\ee
	Expressing the product of epsilons through the Kronecker symbols and properly contracting the indices we obtain: 
	\be
	\left(\tilde{F}F\right)^2 = 2F^4 + 4 F^{\alpha\beta}F_{\nu\alpha}F^{\mu\nu}F_{\mu\beta},
	\label{F_dual_F_sq}
	\ee  
	where $F^4 \equiv (F^{\mn}F_{\mn})^2$.
	We can verify result (\ref{F_dual_F_sq})  expressing the Maxwell tensor through
	electromagnetic fields \textbf{B} and \textbf{E} coming to the well known relation
	$(\tilde{F}F)^2 = \left[-4 \, (\mathbf{E \cdot B})\right]^2$.
	
	The first part of the action (\ref{A-HE-curv}), proportional to $F^4$, 
	leads to the following contribution to the energy-momentum tensor
	\be
	T_{\mn}^{(1)} = C\left(-F^4 g_{\mn}+8F^2 F_{\mu\alpha}F_\nu^{.\,\alpha}\right).
	\label{T_mu_nu_1}
	\ee
	The same contribution, up to a numerical factor, comes from  the first term in eq. (\ref{F_dual_F_sq}), 
	So to find the total EMT
	we need to find the variation of the second term of eq. (\ref{F_dual_F_sq}). Eventually the 
	remaining part of the energy-momentum tensor is
	\be
	T_{\mn}^{(2)}
	=\frac{7C}{4}  \left[ \left( 16 F^2F_{\mu\alpha}F_\nu^{.\,\alpha} - 2F^4 g_{\mn}\right) + 
	\left(32 F_{\mu}^{. \, \lambda}F_{\lambda\alpha}F^{\alpha\beta}F_{\nu\beta} 
	-4F^{\alpha\beta}F_{\lambda\alpha}F^{\sigma\lambda}F_{\sigma\beta} \, g_{\mn} \right)\right].
	\label{T_mu_nu_2}
	\ee
	Bringing together eqs. (\ref{T_mu_nu_1}) and (\ref{T_mu_nu_2}) and raising one index we obtain 
	for the total energy-momentum tensor, originating from the Heisenberg-Euler action, the following  expression:
	\be
	T^{\mu}_{\nu} = -\delta^{\mu}_{\nu}C\left[\frac{9}{2}F^4+
	7F^{\alpha\beta}F_{\lambda\alpha}F^{\sigma\lambda}F_{\sigma\beta}\right] + 
	36CF^2F^{\mu\alpha}F_{\nu\alpha} + 56CF^{\mu\lambda}F_{\lambda\alpha}F^{\alpha\beta}F_{\nu\beta}.
	\label{EMT_EH}
	\ee
	
	Now let us check the conservation of the obtained energy-momentum tensor (\ref{EMT_EH}) in the case of
	constant  $C(T)$. We consider $T_{\mn}^{(1)}$ and $T_{\mn}^{(2)}$ separately.
	\be
	T^{(1)\mu}_{\,\,\,\, \,\,\,\, \nu ;\mu}=-4CF^2F^{\alpha\beta}F_{\alpha\beta ;\nu}
	+8CF^2F^{\mu\alpha}F_{\nu\alpha ;\mu}+8C(F^2F^{\mu\alpha})_{;\mu}F_{\nu\alpha},
	\ee
	where semicolons mean covariant derivatives in the background metric. 
	The last term in this equation is zero according to the equation of motion {corresponding to the Lagrangian
		${\cal L} = F^4$.}

	The first two ones can be rewritten using the relation
	\be
	F_{\alpha\beta;\sigma}+F_{\beta\sigma;\alpha}+F_{\sigma\alpha;\beta} = 0.
	\label{propertyF}
	\ee
	Renaming some {dummy} indices we come to
	\be
	T^{(1)\mu}_{\,\,\,\, \,\,\,\, \nu ;\mu}=4CF^2(F_{\alpha\mu;\nu}+2F_{\nu\alpha;\mu})F^{\mu\alpha}=
	4CF^2(F_{\nu\alpha;\mu}-F_{\mu\nu;\alpha})F^{\mu\alpha}=0.
	\ee
	Here we proved EMT conservation law for those parts of the action which contain $F^4$. It must be analogous 
	for the EMT part originated from $F^{\alpha\beta}F_{\nu\alpha}F^{\mu\nu}F_{\mu\beta}$ 
	(see eq.~(\ref{F_dual_F_sq})).
	\be
	4[-F^{\alpha\beta}F_{\lambda\alpha}F^{\sigma\lambda}F_{\sigma\beta} \, \delta^{\mu}_{\nu} + 8 F^{\mu \lambda}F_{\lambda\alpha}F^{\alpha\beta}F_{\nu\beta}]_{;\mu}=\nonumber\\
	=4[-4F^{\alpha\beta}F_{\lambda\alpha}F^{\mu\lambda}F_{\mu\beta;\nu}+8F^{\mu \lambda}(F_{\lambda\alpha}F^{\alpha\beta}F_{\nu\beta})_{;\mu}],
	\ee
	where we used the Maxwell equation $F^{\mu\nu}_{\,\,\,\,\,\,;\mu} = 0$. 
	Considering the part inside square brackets and taking into account the equation of motion $F^{\mu\lambda}(F^{\alpha\beta}
	F_{\lambda\alpha})_{;\mu} = 0$, that follows from the part of the action: 
	\be
	\mathcal{A}_{HE}^{part}= 7C \int d^4x \, \sqrt{-g} \,  F^{\alpha\beta}F_{\nu\alpha}F^{\mu\nu}F_{\mu\beta}, 
	\ee
	(see eq. (\ref{A-HE-curv}) and (\ref{F_dual_F_sq})), we arrive to 
	\be
	[-4F^{\alpha\beta}F_{\lambda\alpha}F^{\mu\lambda}F_{\mu\beta;\nu}+8F^{\mu \lambda}(F_{\lambda\alpha}F^{\alpha\beta}F_{\nu\beta})_{;\mu}]&=&4F^{\mu\lambda}F^{\alpha\beta}F_{\lambda\alpha}(-F_{\mu\beta;\nu}+2F_{\nu\beta;\mu}) \nonumber\\
	&=&4F^{\mu\lambda}F^{\alpha\beta}F_{\lambda\alpha}(-F_{\mu\beta;\nu}-F_{\beta\mu;\nu}-F_{\mu\nu;\beta} + F_{\nu\beta;\mu})
	\nonumber\\
	&=&4F^{\mu\lambda}F^{\alpha\beta}F_{\lambda\alpha}(-F_{\mu\nu;\beta} + F_{\nu\beta;\mu})=0.
	\ee
	For the transition to the third term of these equalities  we used eq. (\ref{propertyF}).
	
	The conclusion for this section is that EMT originated from 
	the Heisenberg-Euler action with $C = const$ is conserved
	\be
	T^{(HE) \mu}_{\nu;\mu} = 0.
	\ee
	It is noteworthy that
	EMT (\ref{EMT_EH}) is not traceless. Indeed it is equal to
	\be
	T^{\mu}_{\mu}=C\left(18 \, F^4 + 28 \, F^{\mu\lambda}F_{\lambda\alpha}F^{\alpha\beta}F_{\mu\beta}\right).
	\label{EMT_trace}
	\ee

	\subsubsection{Corrections to EMT emerging from the HE action  \label{ss-1order-exp} }

	Now making the usual perturbation expansion (\ref{bar-g}) we find the following first order correction to
	the energy-momentum tensor:
	\be
	T_{\mu\nu}^{HE\,1} = C(T) \left\{ - h_{\mu\nu} \left[(F_{\alpha\beta}  F^{\alpha\beta})^2 -
	\frac{7}{4}({\tilde { F}}^{\alpha\beta}  F_{\alpha\beta})^2 \right] \right. \nonumber\\
	\left. - g_{\mu\nu} \left[ 4 F^2 F_{\alpha\beta} f^{\alpha\beta}  - 
	4 F^2 F_{\alpha\beta} F_{\lambda\sigma} g^{\alpha\lambda} h^{\beta\sigma}
	- \frac{7}{2}( \tilde F  F) {\tilde { F}}^{\alpha\beta}  f_{\alpha\beta}
	\right]  \nonumber \right.\\
	\left.
	+8  F^2  (F_{\mu\lambda} f_\nu^{.\,\lambda} +f_{\mu\lambda} F_\nu^{.\,\lambda})
	+16 F_{\mu\lambda}  F_\nu^{.\,\lambda} F^{\alpha\beta} f_{\alpha\beta}  \right. \nonumber \\
	\left.
	- 8h^{\lambda\sigma} F_{\mu\lambda}  F_{\nu\sigma} F^2
	-16 F_{\mu\lambda} F_\nu^{.\,\lambda} F^{\sigma}_{.\,\beta} F_{\sigma\gamma} 
	h^{\beta\gamma}
	\right\}.
	\label{T-mu-nu-HE1}
	\ee
	This expression can be simplified, because in the absence of background electric field
	$\tilde F F = 0$ and we get:
	\be
	&&T_{\mu\nu}^{HE\,1} = C(T) \left[ - h_{\mu\nu}  (F^2)^2 
	- 4 F^2 g_{\mu\nu} \left( Ff - FFh \right) +  
	8  F^2  (f_{\nu\lambda} F_\mu^{.\,\lambda}
	+f_{\mu\lambda} F_\nu^{.\,\lambda})
	\nonumber \right.\\
	&&+16 F_{\mu\lambda}  F_\nu^{.\,\lambda} (Ff)
	\left.
	- 8h^{\lambda\sigma} F_{\mu\lambda}  F_{\nu\sigma} F^2
	- 16 F_{\mu\lambda} F_\nu^{.\,\lambda} (FF)h
	\right].
	\label{T-mu-nu-HE1}
	\ee
	The trace of this expression is nonvanishing:
	\be
	T_{\alpha}^{\alpha\,HE\,1} = C(T) \left[ 16F^2 (Ff) - 8 F^2 (hFF)-F^4 h
	\right]
	\label{trace-T-HE1}
	\ee
	
		{It is usually demanded in FLRW space-time that the source term $T^{(1)\mu}_{\nu}$ for gravitational wave 
		equation (\ref{GWsys1}) must be traceless.
		To this end one may separate the traceless part out of Eq.~(\ref{T-mu-nu-HE1}) subtracting 
		$ g_{\mu\nu} T_{\alpha}^{\alpha\,HE\,1} /4 $  out of it. However, this prescription would break the conservation of 
		the source and, as is shown in Sec.6 of paper~\cite{OurWork}, it would lead to a violation of  the transversality conditions 
		$D_\mu \psi^\mu_\nu = 0$. Indeed, in ~\cite{OurWork}  we used the condition 
		$\overline{D}_{\mu}\overline{T}^{\mu}_{\nu}=0$ to prove a compatibility of the Einstein equations in the first 
		perturbation order with gauge fixing conditions (\ref{D-psi}).}
	
	Note that the energy-momentum tensor (\ref{T-HE}) is non-conserved for a non-constant $C(T)$
	because the dependence on the temperature appears due to interaction with external system.  So anyhow EMT is not formally
	conserved.
	
	\subsubsection{Summary}
	As conclusion of this subsection, we write the result for the correction to the EMT 
	from the Maxwell action and from the Heisenberg-Euler action, respectively:
	\begin{align}
	\label{EMTMax}
	&T^{Maxwell(1)}_{\mu\nu}=\frac{1}{2}g_{\mu\nu}\left[Ff-FFh\right]+h_{\mu\sigma}F^{\sigma\alpha}F_{\nu \alpha}+h^{\alpha\sigma}F_{\mu\alpha}F_{\nu\sigma}-f_{\mu\alpha}F_{\nu}^{.\,\alpha}-F_{\mu\alpha}f_{\nu}^{.\,\alpha},\\
	\label{EMTHE}
	&T_{\mu\nu}^{HE\,(1)} = C(T) \left[ - h_{\mu\nu}  (F^2)^2 - 4 F^2 g_{\mu\nu} \left( Ff - FFh \right) +  8  F^2  (f_{\nu\lambda} F_\mu^{.\,\lambda}+f_{\mu\lambda} F_\nu^{.\,\lambda})\right]\nonumber\\
	&+C(T)\left[16 F_{\mu\lambda}  F_\nu^{.\,\lambda} (Ff) - 8h^{\lambda\sigma} F_{\mu\lambda}  F_{\nu\sigma} F^2 -16 F_{\mu\lambda} F_\nu^{.\,\lambda} (FFh)\right].
	\end{align}
	where $Ff=F^{\alpha\beta}f_{\alpha\beta},\,\,\,
	FFh=h^{\alpha}_{\sigma}F_{\alpha\beta}F^{\sigma\beta}$.
		
	\subsection{Maxwell tensor and cosmic magnetic and electric fields \label{ss-maxwell-B}}
	
	Eqs. (\ref{EMTMax}-\ref{EMTHE}) look {quite}
	complicated. Further we simplify this equations and express them in terms of physical magnetic field. 
	To understand the physical meaning of the different components of $F_{\mu\nu}$, $F^{\mu\nu}$, 
	or $F_{\mu}^{\nu}$, let us start from the geodesic equation for a charged particle in external electric and
	magnetic field (see e.g. book~\cite{LL-2}. Eq.~(90.7)):
	\be
	m \frac{D u^\alpha}{ds} = e F^\alpha_{.\,\,\beta} u^\beta,
	\label{Du-ds}
	\ee
	where $u^\alpha = dx^\alpha/ds$ is the particle four-velocity. From this equation it is clear that
	physical electric field is the Maxwell tensor with mixed components, $E^j = F_0^j $, and physical magnetic field is 
	expressed through the Maxwell tensor  $F^i_j$ as:
	\be 
	B_1 = F^2_{.\,3},\,\,\,  B_2 = -F^1_{.\,3}, \,\,\, B_3 = F^1_{.\,2},
	\label{B-j}
	\ee 
	or in compact form $B_i = \epsilon_{ijl} F^j_{.\, m} \delta^{m l} $.
		
	The first pair of Maxwell equation has the same form as in flat space-time:
	\be
	\partial_\lambda F_{\mu\nu} + \partial_\nu F_{\lambda \mu}+\partial_\mu F_{\nu\lambda} = 0.
	\label{max-eq-1}
	\ee
	If the background electric field is absent, i.e. $F_{tj} = 0$, then
	\be 
	\partial_t F_{ij} = 0. 
	\label{dt-F-ij}
	\ee
	Hence $F_{ij}$ remains constant in the process of cosmological expansion and correspondingly physical magnetic 
	field behaves as:
	\be
	F_i^{.\,j} = g^{j k} F_{i k} = - F_{jk} / a^2.  
	\label{B-phys}
	\ee
	In other words, physical magnetic field drops as $1/a^2$, the well known result.
	
	If electric field is absent  and only external magnetic field is non-zero, then
	the dual Maxwell tensor (\ref{tilde-F-c}) has only space-time components. The quantity
	$D_{tj} =\tilde F_{tj} /\sqrt{-g} = (1/2) \epsilon_{tjlm}F^{lm} $ is expressed through magnetic field as
		\be
		D_{tj} = - D_{jt} = B_j /a^4 .
	\label{D-tj}
	\ee 
	
	In flat space-time varying magnetic field induces electric field according to
	\be
	\nabla \times E = -\partial_t B. 
	\label{rot-E}
	\ee 
	In curved space-time the analogue of this equation is Eq.~(\ref{max-eq-1}) with $\lambda =t$ or
	Eq.~(\ref{dt-F-ij}), so if originally electric field was absent, it would not be induced by time-varying magnetic field,
	in the case that the time variation is created by the cosmological expansion (\ref{B-phys}).
	
	In terms of physical magnetic field $\bf{ B}$ the product $F_{\mu\nu} F^{\mu\nu}$ with indices lifted by the
	background metric $g^{\mu\nu}$ is
	\be
	F_{\mu\nu} F^{\mu\nu} = F^2 =  2 {\bf B}^2 \sim 1/a^4.
	\label{FF}
	\ee

	\subsection{Scalar and tensor mode mixing in external magnetic field}
	
	Using equations (\ref{Ricci}-\ref{hii}), we rewrite equation (\ref{EOMtrace}) for the case of the FLRW metric as:
	\be\label{EOMtraceFLRW}
	\partial^2 h-12\frac{\ddot{a}}{a}h_{00}-4\left(\frac{\ddot{a}}{a}+2H^2\right)h^i_i+6\left(\frac{\ddot{a}}{a}+H^2\right)h
	=16\pi G T^{\alpha(1)}_{\alpha} .
	\ee

	The EMT perturbation originating from the Maxwell action is traceless, while that from the Heisenberg-Euler action 
	has a non-zero trace. {Indeed,}
	\be
	T_{\alpha}^{\alpha\,HE\,1} = C(T) \left[-hF^4+16F^2 (Ff) - 8 F^2 (FFh)\right].
	\label{trace-T-HE1}
	\ee
Now one could naively divide the source into a traceless part and a non-zero trace part (simply subtract the trace multiplied by the 
background metric). {To this end} let's look at Eq.(\ref{EOMtraceFLRW}) {and explicitly} 
substitute $T^{\alpha\, EM(1)}_{\alpha}$ into the right-hand side. We get
	\be\label{EMTcorrectiontrace}
	&\partial^2 h-12\frac{\ddot{a}}{a}h_{00}-4\left(\frac{\ddot{a}}{a}+2H^2\right)h^i_i+6\left(\frac{\ddot{a}}{a}+H^2\right)h=\nonumber\\
	&=16\pi GCF^2\left[-hF^2+16(Ff) - 8 {(F_{\alpha\beta}F^{\alpha\sigma}h^{\beta}_{\sigma})}\right].
	\ee
	{We see} that the equation contains both scalar and tensor parts. {Thus} it is 
	impossible to write a separate equation for each mode.
	To make this even more obvious, let us fix the coordinates so that the magnetic field is directed along the $x$ axis.
	Then the following components of the electromagnetic field tensor will be non-zero:
	\be
	&F^y_{.\,z}&=-F^z_{.\,y}=B_x.\\
	&F^{yz}&=-F^{zy}=-\frac{B_x}{a^2}.\\
	&F_{yz}&=-F_{zy}=-B_x a^2.
	\ee
	The trace from the correction to the EMT can then be rewritten taking into account the following relations:
	\begin{gather}
	F^2=2B^2,\\
	Ff=F^{.\,\beta}_{\alpha}f^{\alpha}_{.\,\beta}=F^{.\,z}_y f^y_{.\,z}+F^{.\,y}_z f^z_{.\,y}=B\left(f^y_{.\,z}-f^z_{.\,y}\right)=2Bf^y_{.\,z}\label{Ff},\\
	FFh=h^{\alpha}_{\sigma}F_{\beta\alpha}F^{\beta\sigma}=B^2\left(h^y_y+h^z_z\right)\label{FFh}.
	\end{gather}
	And in equation (\ref{EMTcorrectiontrace}) we have
	\be\label{EMTcorrectiontraceExplicit}
	&\partial^2 h-12\frac{\ddot{a}}{a}h_{00}-4\left(\frac{\ddot{a}}{a}+2H^2\right)h^i_i+6\left(\frac{\ddot{a}}{a}+H^2\right)h=\nonumber\\
	&=32\pi GCB^3\left[-hB+16f^y_{.\,\,z} - 4 {B\left(h^y_y+h^z_z\right)}\right].
	\ee
	The diagonal components can be written as the sum of scalar and tensor quantities of the helicity expansion
	\be
	h^y_y=-2\Psi \delta^y_y +D^y_y,\\
	h^z_z=-2\Psi \delta^z_z +D^z_z.
	\ee
	And, substituting the helicity expansion into the complete equation, we obtain the final equation, which shows the mixing 
	of scalars $\Phi, \Psi$ and tensor $D_{ij}$:
	\be
	&\left(\partial^2+64\pi G CB^4\right) \left(\Phi+3\Psi\right)+6\left(\Psi-\Phi\right)\left(\frac{\ddot{a}}{a}-H^2\right)=\nonumber\\
	&=32\pi GCB^3\left[16f^y_{.\,\,z} - 4 {B\left(4\Psi+D^y_y+D^z_z\right)}\right],
	\ee
	where $h=2\Phi+6\Psi$. 
	
	In addition, we note that there may also be implicit mixing through the term with $f^y_{.\,\,z}$ {in the above equation,} 
	since the equation for the electromagnetic wave contains various convolutions of tensors 
        with tensor $h_{\mu\nu}$ ({ see below, eq. (\ref{d2-f})).}
	
As was noted in the Sec.\ref{Mixing}, the result is quite evident, since the external magnetic field gives, in addition to the GW propagation vector, another preferred direction in space. This leads to the loss of axial symmetry and to mixing of the 
scalar and tensor modes of metric perturbations.
	
	\section{Electromagnetic wave propagation in external magnetic field}\label{EMWprop}
	
	In this section, we will derive the equation for the propagation of electromagnetic waves in curved space-time and in the presence 
	of an external magnetic field, thereby completing the derivation of the system of differential equations (SoDE) for the metric-EMF 
	perturbation system. We will briefly call this system $g-\gamma$, by $g$ 
	we mean a graviton with any possible polarization - 0, 1, 2.
	
	\subsection{Equation of motion from the Maxwell action}	
	
	Variation of the Maxwell action from eq.(\ref{A-Max}) over $ \delta A_\nu$ leads to the equation of motion 
	$\bar D_\mu \bar F^{\mu\nu} = \bar J^\nu$, where $\bar D_\mu$ is the covariant derivative in the full metric
	$\bar g_{\mu\nu}$. Due to antisymmetry of $\bar F^{\mu\nu} $ this equation is reduced to: 
	\be 
	\bar D_\mu \bar F^{\mu\nu}= \frac{1}{\sqrt{-\bar g}} \partial_\mu \left(   \bar F^{\mu\nu} \sqrt{-\bar g} \right) = \bar J^\nu
	\label{d-F-Max}
	\ee
	Below we assume that neither electric charge density nor electric current are present, i.e.
	$\bar J^\nu= 0$.

	Substituting expansions (\ref{bar-g}), (\ref{def-F-f}), and (\ref{det-g-expan}) 
	into Eq. (\ref{d-F-Max}) we obtain
	\be
	&\frac{1}{\sqrt{-\bar g}}&\partial_\alpha\left(\sqrt{- \bar g}\bar{F}^{\alpha\beta}\right)=\frac{1}{\sqrt{- \bar g}}\partial_\alpha\left(\sqrt{- \bar g}\,\bar{g}^{\alpha\alpha'}\,\bar{g}^{\beta\beta'}\bar{F}_{\alpha'\beta'}\right)=
	\nonumber\\
	&\frac{1}{\sqrt{- \bar g}}&\partial_\alpha\left[\sqrt{-\bar g}\,\left(g^{\alpha\alpha'} - h^{\alpha\alpha'}\right)\,\left(g^{\beta\beta'} - h^{\beta\beta'}\right)\left(F_{\alpha'\beta'} + f_{\alpha'\beta'}\right)\right] = 0.
	\label{eom_maxwell_pert}
	\ee
	The external electric field is supposed to be zero and only background magnetic field is present, so $F_{t\beta} = 0$.
	Thus the zero order term, which is the equation of motion for the
	background magnetic field, has the form;
	\be
	\partial_\mu  F^{\mu\nu}  =    
	\partial_\mu \left( g^{\mu \alpha} g^{\nu \beta} F_{\alpha\beta} \right)= 0.
	\label{div-B}
	\ee
	This is the analogue of the equation $div\, {\bf B} =0$ in flat space-time.

	In FLRW metric the metric determinant is expanded as:
	\be 
	\sqrt {-\bar g} = \sqrt {- g} (1+ h/2) = a^3 (1+h/2).
	\label{det-g-2}
	\ee
	and so equation (\ref{eom_maxwell_pert}) takes the form:
	\be \partial_\mu \left[ \left( g^{\mu\alpha} -  h^{\mu\alpha}\right) \left( g^{\nu\beta} -  h^{\nu\beta}\right)
	\left( F_{\alpha\beta} + f_{\alpha\beta} \right) \right] + F^{\mu\nu} \partial_\mu h/2
	+ 3H  g^{\nu\beta}  f_{ t \beta} = 0 ,   
	\label{d-F-2}
	\ee
	where we took into account that $g^{t\alpha} = \delta^{t\alpha}$ and $F_{t\beta} = 0$.
	We also  assume that $f_t = f^t =0$ and
	impose the transversality condition on the propagating photon modes:
	\be
	{D_\mu f^\mu = \frac{1}{\sqrt{-g}} \partial_\mu \left(\sqrt{-g} f^\mu \right) =
		\partial_t f^t + 3 H f^t - \partial_j f^j = 0 ,
		\label{d0f}}
	\ee
	which for $f^t = 0$ leads to $\partial_j f^j = 0$.

	Thus the first order expansion of eq.~(\ref{d-F-2}) has the form:
	\be
	\partial_\mu f^{\mu\nu} + 3H f^{t\nu} - \partial_\mu \left(h^\mu_\alpha F^{\alpha\nu} + 
	h^\nu_\alpha  F^{\mu\alpha} \right)  + F^{\mu\nu} \partial_\mu h/2 + Q^\nu = 0 .
	\label{d-mu-fmu}
	\ee
	where we introduced a new quantity $Q^\nu$ to describe contribution from different additional terms such as 
	Heisenberg-Euler corrections, matter effects, etc to be considered below.

	To derive the first order equation for $f_j$ we multiply eq. ~(\ref{d-mu-fmu})
	by $g_{\nu j} = -a^2 \delta_{\nu j}$ (latin indices are always supposed to be the space ones, 
	e.g. $j= 1,2,3$) and recall that $F_{t\mu} = 0$ and
	$f_t = 0$. So  we get  finally:
	\be 
	\partial_t^2 f_j - \frac{\Delta f_j}{a^2} 
	+ H \partial_t f_j -  g_{\nu j}  \partial_\mu \left(h^{\mu \alpha} g^{\nu\beta} F_{\alpha\beta} + 
	g^{\mu \alpha} h^{\nu\beta} F_{\alpha\beta}  \right)  + F^{\mu}_{.\,j} \partial_\mu h/2
	+ Q_j   = 0 ,
	\label{d2-t-f1}
	\ee
	where  $Q_j =  g_{\nu j}  Q^\nu = -a^2 \delta_{\nu j} Q^\nu $ and
	$\Delta$ is the flat space Laplacian. 
	
	To proceed further we have to fix certain gauge conditions on metric perturbations $h_{\mu\nu}$. 
	We will follow our paper~\cite{OurWork}, where it is shown that the following conditions can be
	imposed in arbitrary background metric:
	\be
	D_\mu\psi^\mu_\nu \equiv D_\mu \left(h^\mu_\nu - h \delta^\mu_\nu/2 \right) = 0,
	\,\,\,\, h_{t\mu} = 0 .
	\label{D-psi}
	\ee
	Since in FLRW metric the only non-zero components of the Christoffel symbols are:
	\be
	\Gamma_{ti}^j = H \delta^j_i, \,\,\,\, \Gamma_{ij}^t = H a^2 \delta_{ij}
	\label{Gamma-ij}
	\ee
	the covariant derivative of $h_j^i$ is reduced to the ordinary derivative and 
	\be
	D_\mu h^\mu_\nu = 
	\partial_\mu h^\mu_\nu  -  \Gamma_{\mu\nu}^\lambda  h^\mu_\lambda +
	\Gamma_{\mu\lambda}^\mu  h^\lambda_\nu =  \partial_\mu h^\mu_\nu.
	\label{D-h-ij}
	\ee
	So using Eq.~(\ref{div-B}), (\ref{D-psi}), and the absence of external electric field, $F_{tj} = 0$, we obtain:
	\be 
	\partial_t^2 f_j - \frac{\Delta f_j}{a^2} 
	+ H \partial_t f_j -    h^{m}_{ i} \partial_m F^{i}_{.j} - 
	F^{m}_{.i} \partial_m h^{i}_{j}  
	+ Q_j  
\     \equiv  {\cal M} [f_j] + Q_j   = 0 .
	\label{d2-t-fn}
	\ee
	The terms proportional to $\partial_ m h $ cancel out  because 
	$F^i_{.j} \partial_m h^m_i = F_{.j}^{i} \partial_i h/2$.
	
	Here we have introduced the new notation:
	\be
	{\cal M}_j [f]  = \partial_t^2 f_j - \frac{\Delta f_j}{a^2} 
	+ H \partial_t f_j -    h^{m}_{ i} \partial_m F^{i}_{.j} - 
	F^{m}_{.i} \partial_m h^{i}_{j} 
	\label{M-fj}
	\ee
	to  be used in what {follows.}
	
	\subsection{Equation of motion from the Heisenberg-Euler action \label{ss-eq-of-mot-HE} }
	The variation of ${\cal A}_{HE}$ (\ref{A-HE-curv}) 
	over $\delta A_\nu$ results in the following contribution to the electromagnetic 
	field equation:
	\be
	{\bar D_\mu \bar F^{\mu}_{. j}}  
	{+ \bar Q^{(HE)}_j = 0,}
	\label{Max-eq-HE}
	\ee
	where {the first term originated from the variation of the Maxwell action (see previous subsection), while the second 
		is the contribution from  $ {\cal A}_{HE} $ (\ref{A-HE-curv}) and has the following form}
	\be 
	\bar Q^{(HE)}_j &=& - g_{\nu j} D_\mu \left[ C(T)  \left( 8 \bar F^{\mu\nu}  \bar F^2 
	+ 14 \tilde{\bar F}^{\mu\nu}  ({\tilde {\bar F}^{\alpha\beta}} \bar F_{\alpha\beta}) 
	\right)\right] = \nonumber\\
	&=&- \frac{g_{\nu j}}{\sqrt{- \bar g}} \partial_\mu \left[ C(T) \sqrt{- \bar g}\left( 8 \bar F^{\mu\nu}  \bar F^2 
	+ 14 \tilde{\bar F}^{\mu\nu}  (F \tilde{F})
	\right)\right] .
	\label{d-F-HE}
	\ee 
	where $F^2=F_{\mu\nu}F^{\mu\nu}$ and $(F \tilde{F}) = { \tilde F_{\mu\nu} F^{\mu\nu}}$.
	
	{We have shown in {Sec.\ref{ss-maxwell-B}}  that the free external magnetic field is not constant, but rises backward in time with the decreasing scale factor as $1/a^2$.}
	
	Let us return to Eq.~(\ref{d-F-HE}) and make perturbative expansion according to Eq.~(\ref{bar-g}). We start from
	consideration of the first term in square brackets, which with account of 
	the  zeroth and first orders  terms takes the form 
	\be
	{g_{\nu j}}\frac{\delta {\cal A}_{HE1}}{\delta \bar A_\nu} =
	g_{\nu j}\frac{(-1)}{\sqrt{-g(1+h)}} \partial_\mu \left\{ 8 \sqrt{-g(1+h)} C \left[ F^2 F^{\mu\nu} + F^2 f^{\mu\nu} +
	2 F^{\mu\nu} (f F) 
	\right.\right. \nonumber\\
	\left.\left. - 2 F^{\mu\nu} {(FFh)}
	 - F^2 \left(h^{\mu}_{\alpha} F^{\alpha\nu} +h^{\nu}_{\alpha}F^{\mu\alpha} 
	\right) \right] \right\} ,\,\,\,\,\,\,\,\,\,\,\,
	\label{HE-EM-1}
	\ee
	where $f F = f_{\mu\nu} F^{\mu\nu}$, $(FFh) = h^{\alpha}_{\sigma} F_{\alpha\beta} F^{\beta \sigma} $, 
	and indices are shifted up or down with the background metric.
	
	The zero order term in this equation somewhat changes the equation of motion of
	the background magnetic field {in FLRW metric} leading to:
	\be 
	\partial_j F^{j \nu} - 16 C \partial_j \left({ B}^2  F^{j \nu} \right) = 0 ,
	\label{d-F-Max-HE}
	\ee
	which is not of much importance for the evolution of $B_j$. 
	The terms proportional to the time derivatives of $\sqrt{-g}$, $C$, and 
	$F^2$ do not appear if $F^{t\mu} = 0$ and $h^t_\alpha =0$.

	The first order part of expression~(\ref{HE-EM-1}) is equal to:
	\be
	Q^{HE1}_j =&-&16 g_{\nu j} C\partial_\mu \left[ B^2 f^{\mu\nu}
	+ F^{\mu\nu}(f F) 
	+ F^{\mu\nu} (FFh)
	-B^2\left(h^{\mu}_{\alpha} F^{\alpha\nu} + h^{\nu}_{\alpha}F^{\mu\alpha} \right)\right]
	- \nonumber\\
	&-&{16Cg_{\nu j}F^{\mu\nu} B^2\partial_{\mu} h /2}
	-16 B^2\left(\dot{C}+3HC\right)f^{t}_{.j} .
	\label{HE-1}
	\ee
	
	The term proportional to $B^2 C$ in this expression,  has the form:
	\be
	{-}16C B^2 g_{\nu j} \left( \partial_\mu f^{\mu\nu} + 3 H f^{t\nu} -
	h^\mu_\alpha \partial_\mu F^{\alpha \nu} -F^{\mu\alpha} \partial_\mu h^{\nu}_{\alpha} 
	\right) \equiv -16CB^2 {\cal M}_j[f] ,
	\label{C-df}
	\ee
	where ${\cal M}_j[f] $ is defined in the equations (\ref{d2-t-fn},\,\ref{M-fj}).
	The factor in the brackets in the l.h.s. of the above equation 
	coincides with the l.h.s. of 
	Eq.~(\ref{d2-t-fn}), except for the last term $Q_j$,
	so it can be absorbed into Eq.~(\ref{d2-t-fn}) changing the overall coefficient from 1 to
	$(1 -16C B^2)$.
	
	{In addition to the terms proportional to $B^2$, 
		the first two terms in Eq.~(\ref{HE-1}) give the following
		contribution of the first part of the HE action to the  photon propagation equation:}
	\be 
	{-}16C g_{\nu j} \left[ f^{\mu\nu} \partial_\mu B^2 {+} F^{\mu\nu} \partial_\mu(f F) \right] 
	{=-16C\left[f^{\mu}_{.j}\partial_{\mu}B^2+F^{\mu}_{.j}\partial_{\mu}(Ff)
	\right] .}
	\label{2-1}
	\ee
	
	{
	So all the terms in  Eq~(\ref{HE-1}), except for those 
	absorbed into Eq.~(\ref{d2-t-fn}) and 
	containing $h_{\mu\nu}$, turn into:
	\be
	Q^{HE11}_j=-16  B^2 {\dot C}f_{t j}-
	16C\left[f^{\mu}_{.j}\partial_{\mu}B^2+F^{\mu}_{.j}\partial_{\mu}(Ff)
	\right].
	\label{2-1-fin}
	\ee	
	The contribution of the terms containing $h_{\mu\nu}$ in
	Eq.~(\ref{HE-1}) can be written as
	\be
	Q^{HEh}_j=-16 C \left[  F_{.j}^{\mu} \partial_{\mu}(FFh) - 
	(h^{\mu\alpha}  F_{\alpha j} +h_{\alpha j} F^{\mu\alpha})\partial_{\mu} B^2
	\right] .
	\label{HE-h}
	\ee
	Finally, for the total $Q_j^{HE1}$ we obtain
	\be 
	Q^{HE1}_j&=&-16CB^2 {\cal M}_j[f]+Q_j^{HE11}+Q_j^{HEh}=\nonumber \\
	&=&-16C\left[B^2\left({\cal M}_j[f] +\frac{\dot{C}}{C}f_{tj}\right)+\left(f^{\mu}_{. j}-
	h^{\mu\alpha}F_{\alpha j}-h_{j \alpha}F^{\mu\alpha}\right)\partial_{\mu}B^2\right]-\nonumber \\
	&-&16CF^{\mu}_{. j} \partial_{\mu}\left(fF+FFh\right),
	\label{Q-HE1}
	\ee
	where ${ \cal M}_j [f]$ is defined in eq.~(\ref{M-fj}).
	
	It is convenient to introduce auxiliary vector field through the equation
	\be 
	F_{ij} = -\epsilon_{ijk} {\cal H}^k 
	\label{H-k}
	\ee
	Physical magnetic field is related to $\cal H$ as $B_k = {- {\cal H }^k}/ a^2={- {\cal H }_k a^2}$ 
	{(because $B_k \sim F^i_{.j} \sim g^{il}F_{lj} \sim \delta_{il}F_{lj} /a^2  $)}.
	
	To decipher the last term in Eq.~(\ref{Q-HE1}) we use the identity:
	\be
	{
		\epsilon_{dbc} \epsilon^{ljm} = \delta_{d}^{l} \delta_{b}^{j} \delta_{c}^{m}+
		\delta_{d}^{m} \delta_{b}^{l} \delta_{c}^{j}+
		\delta_{d}^{j} \delta_{b}^{m} \delta_{c}^{l}-\delta_{d}^{j} \delta_{b}^{l} \delta_{c}^{m}-
		\delta_{d}^{m} \delta_{b}^{j} \delta_{c}^{l}-\delta_{d}^{l} \delta_{b}^{m} \delta_{c}^{j}.
	}
	\label{eps-2}
	\ee
	and Eq.~(\ref{H-k}). So $FFh$ and $F^{\mu}_{.j}\partial_{\mu}(fF)$ turn into
	\be
	FFh=F_{\alpha\beta}F^{\beta\sigma}h^{\alpha}_{\sigma}=-B^2 h^i_i+h^{ij} B_i B_j,
	\label{FFh__}
	\ee
	\be
	F^{\mu}_{.j}\partial_{\mu}(fF)&=&F^{m}_{.j}\partial_{m}(fF)=2g_{nj}B_k \partial_{m}\left( f^{mn}B_k+f^{km}B_n+f^{nk}B_m\right)\nonumber \\
	&=&2B_k \partial_{m}\left( f^{m}_{.\,j}B_k+f^{km}B_j+f_{j}^{.\,k}B_m\right),
	\label{FdfF__}
	\ee
	where summation {over repeated indices is performed.}
	}

	The variation of the second term in the HE action, Eq.~(\ref{d-F-HE}),
	is equal to:
	\be
	\frac{\delta {\cal A}_{HE2}}{\delta \bar A_\nu} =\frac{(-14)}{\sqrt{- \bar g}} \partial_\mu \left[ \sqrt{- \bar g} \,C(T)
	\tilde{\bar F}^{\mu\nu}  ({\tilde {\bar F}^{\alpha\beta}} \bar F_{\alpha\beta}) 
	\right] .
	\label{d-F-HE-2}
	\ee
	In the case that the background electric field is absent and only magnetic field is non-zero,
	the r.h.s. of the equation above vanishes in the zeroth perturbation order  because $F_{\alpha\beta}$ is non-zero
	only for space-space components, while $ \tilde F_{\alpha\beta}$ is non-zero
	for space-time components. Hence  $ \tilde F_{\alpha\beta}  F^{\alpha\beta} =0$.

	{Accordingly, expression (\ref{d-F-HE-2}) multiplied by $g_{j\nu} $ can be expanded as
	\be
	Q^{HE2}_j=-28\left[(\dot{C}+HC) \delta_\mu^t+C\partial_{\mu}\right]\tilde{F}^{\mu}_{. j}(f\tilde{F}),
	\label{d-F-HE-3}
	\ee
	where 
	\be
	\tilde{F}^{\mu}_{. j}(f\tilde{F}) =g_{\nu j}\tilde{F}^{\mu\nu}(f\tilde{F})=\frac{1}{4} g_{\nu j}\epsilon^{\mu\nu\alpha\beta} F_{\alpha\beta}\,
	\epsilon_{\sigma\lambda\rho\gamma} F^{\sigma\lambda} f^{\rho\gamma}. 
	\ee
	So using Eq.~(\ref{H-k}) and contracting
	\be
	\epsilon^{jkl}\epsilon_{mkl} = 2 \delta^j_m
	\label{eps-jkl}
	\ee
	we obtain from Eq.(\ref{d-F-HE-3}):
	\be
	Q_j^{HE2} =-
	56 C \left[ \left(\frac{\dot C}{C} - 3 H \right) B_j B_k \partial_t f_k + B_j B_k \partial^2_t f_k
	\right],
	\label{HE2-2}
	\ee
      where the summation over the repeated space indices is made with Kronecker delta and
	it is taken into account that $B_j \sim 1/a^2$.
	
	So using Eqs.~(\ref{d2-t-fn},\,\ref{M-fj}\,,\ref{Q-HE1},\,\ref{d-F-HE-3})  we obtain:
	\be 
	\partial_t^2 f_j - \frac{\Delta f_j}{a^2} 
	+ H \partial_t f_j -    h^{m}_{ i} \partial_m F^{i}_{.j} - 
	F^{m}_{.i} \partial_m h^{i}_{j} 
	+Q^{HE1}_j+Q^{HE2}_j
	\equiv \nonumber\\
	{\cal M}_j [f] +Q^{HE1}_j+Q^{HE2}_j
	=0
	\label{fj-fin}
	\ee
	and come to the almost final equation for photons
	\be
	\left( 1 - 16 CB^2 \right){\cal M}_j [f]-16B^2\dot{C}f_{tj}-16{C}\left(f^{\mu}_{. j}-h^{\mu\alpha}F_{\alpha j}-h_{j \alpha}F^{\mu\alpha}\right)\partial_{\mu}B^2-\nonumber \\
	-16CF^{\mu}_{. j} \partial_{\mu}\left(fF+FFh\right)-
	28\left[(\dot{C}+HC){\delta^{t}_\mu}+C\partial_{\mu}\right]\tilde{F}^{\mu}_{. j}(f\tilde{F})=0 .
	\label{almost-fin}
	\ee
	We remind that   ${\cal M}_j [f] $ is defined in eq.~(\ref{M-fj}).

	
	\subsection{Conformal anomaly effect \label{ss-conf-anom}}
	
	Quantum corrections  to the energy-momentum tensor of electromagnetic field $T_{\mu\nu}^{(em)}$
	in curved space-time background lead to the
	the well known conformal anomaly, for a review see ref.~\cite{conf-anom}, resulting in the nonzero trace of
	the electromagnetic energy-momentum tensor:
	\be
	T_\mu^{\mu(anom)} = \frac{\alpha\beta}{8\pi}\,G_{\mu\nu}^{(a)} G^{\mu\nu (a)}
	\label{T-mu-mu-anom} 
	\ee
	where $G_{\mu\nu}$ is the gauge field stress tensor, $\alpha $ is the fine structure constant and $\beta$ is the first coefficient of the beta-function expansion for 
	the gauge group of rank $N$:
	\be
	\beta = \frac{11}{3} N - \frac{2}{3} N_F,
	\label{beta}
	\ee
	with $N_F$ being the number of the fermion species.
	
	There are additional contributions into the trace proportional to the products of the Riemann, Ricci tensors, and curvature scalar
	which are generally nonlocal~\cite{Bunch-Davies,Birrell-Davies}. We will not consider them in this work.
	
	The trace anomaly allows for photon production by the conformally flat gravitational field~\cite{AD-anom, AD-anom-2}
	in contrast to the Parker theorem~\cite{Parker}.
	
	The Fourier transform of the amplitude of the photon propagation in the gravitational field has pole at $q^2 =0$,
	where $q$ is the four-momentum transfer to gravitational field. According to the result of paper~\cite{AD-anom} 
	the anomalous part of the energy-momentum tensor has the form:
	\be
	T_{\mu\nu}^{(anom)} \sim \frac{q_\mu q_\nu - g_{\mu\nu} q^2}{q^2} F_{\alpha\beta} F^{\alpha\beta}.
	\label{T-anom}
	\ee
	It is evidently conserved and has non-zero trace.
	
	As is shown in Ref.~\cite{AD-anom}, conformal anomaly (\ref{T-mu-mu-anom}) 
	leads to an additional contribution to eq.~(\ref{d2-t-f1}) or, which is essentially the same, to eq.~(\ref{d2-t-fn}).
	\be
	\alpha \beta \left( \partial_\mu F^\mu_{. \nu}  \ln a - H F_{.\nu}^{t} \right).
	\label{anom-eq} 
	\ee
	The first term here is the usual charge renormalization and the second one is the anomaly 
	giving rise to photon production in conformally flat metric. This metric allows for 
	the transformation to the conformal time leading to {the Minkowski} metric proportional 
	to the common scale factor. The canonical Maxwell equation, without the anomalous term, transforms in this metic 
	into the free Maxwell equation in flat space-time, while the additional anomalous term does not allow that.
	
	\subsection{Plasma interaction effects\label{ss-plasma}}
	
	Photons propagating in the primeval plasma interact with the plasma particles and as a result they 
	acquire an effective mass, the so called plasma frequency, $\Omega_{pl} $, so the relation between 
	photon frequency, $\omega$, and momentum, $k$, changes as $\omega^2 - k^2 = \Omega_{pl}^2$.
	Waves with $\omega < \Omega_{pl} $ do not propagate in plasma. 
	
	{In the canonic theory the 
	effective action describing the plasma frequency term is usually written as}
	\be
	\mathcal{A}_{pl} = \frac{1}{2} \int d^4 x \sqrt{-g}\, \Omega_{pl}^2\, g_{\mu\nu}\, f^\mu f^\nu .
	\label{A-plasma} 
	\ee
	{This term is proportional to the square of the small amplitude $f^\mu$ of the electromagnetic wave and seemingly
	should be neglected in our first order approximation. However, this is not so because to obtain the first order equation
	one has to take the action in the second order in small quantities. The first order terms are absent in the action since
	$f^\mu$ satisfies the equation of motion that are realised at the extremal value of action for which 
	$\delta A /\delta f^\mu = 0$.}
	
	The corresponding energy-momentum tensor is
	\be
	T_{\mu\nu} \sim f^2
	\label{T-mu-nu-pl}
	\ee
	is quadratic in $f$ and can be disregarded in our approximation.{It is similar to a scalar field with small
	amplitude $\phi$ that has energy density proportional to $m_\phi^2 \phi^2$, so its energy-momentum tensor
	is quadratically small but non-zero mass, $m_\phi$ is essential for propagation of $\phi$ waves.}

	All electrically charged particles contribute
	to plasma frequency. If the particle mass is larger than the temperature of the relativistic cosmological plasma, 
	$m >T$, the contribution to plasma frequency from such nonrelativistic charged particles is
	\be
	\Omega_{pl}^{nr} = \frac{ e^2 n}{m},
	\label{omega-nr}
	\ee
	where $n$ is the number density of particles with charge $e$, $e^2 = 4\pi \alpha$.
	Note that the number density in this case is exponentially suppressed, $n\sim \exp(-m/T)${\cite{Pitaevskii}.}
	
	{On the other hand} relativistic particles, with $m<T$, contribute as:
	\be
	\Omega_{pl}^{rel} = \frac{ 2T^2 }{9} \,\sum_j  e_j^2 ,
	\label{omega-rel} 
	\ee
	where the summation is done over all relativistic charged particles with charges $e_j$. 
	The electric charge, $e$, depends on temperature due radiative corrections{\cite{RelPlasma}}. 
	
	Plasma frequency is determined
	by the photon Green's function in the limit of vanishing photon momentum. More rigorous treatment of the
	problem of the photon propagation in plasma demands determination of the proper Green's function.
	Simple derivation of these expression including the Green's function can be found in ref.~\cite{Lepidi}. However,
	in this paper we will use simplified approximation describing the plasma effects by the plasma frequency only.
	
	We need also take into account the loss of coherence of the photons produced by gravitons. We
	describe this phenomenon  introducing damping term into equation of motion for photons in the
	form $\Gamma \dot f_j$, where we approximate $ \Gamma$ as
	\be 
	\Gamma = v \sigma n,
	\label{Gamma}
	\ee
	where $n = 0.1 g_* T^3$ is the density of charged particles in plasma, $g_*= 10-100$ is the number of
	charged particle species, $v \sim 1 $ is the 
	relative velocity of "our" photon
	and the scatterer in plasma, and $\sigma = \alpha^2/T^2$ is the scattering cross-section.
	
	So the final equation for photons propagating in an arbitrary curved space-time background and external 
	magnetic field in cosmic plasma with an account of the photon collisions with plasma particles has the form:
	\be
	&& \left( 1 -  16 C B^2\right) \left[ \partial_t^2 f_j - \frac{\Delta f_j}{a^2} + \frac{\partial_j \partial_k f_k}{a^2}
	+ H \partial_t f_j -  g_{\nu j}   \left(h^{\mu \alpha}  \partial_\mu  F^\nu_{.\alpha} + 
	F^\mu_{. \alpha} \partial_\mu h^{\nu\alpha}
	\right)  \right]  \nonumber \\
	&& -16 B^2 \dot C f_{tj}
	- 16 C \left[ \left(f^{\mu}_{. j} -
	h^{\mu\alpha}F_{\alpha j}-h_{j \alpha}F^{\mu\alpha}\right)\partial_{\mu}B^2 
	+ F^{\mu}_{. j} \partial_{\mu}\left(fF-FFh\right)\right] \nonumber \\
	&& -28\left[(\dot{C}+HC)\delta^t_{\mu}+C\partial_{\mu}\right]\tilde{F}^{\mu}_{. j}(f\tilde{F}) +
	\alpha \beta \left[  \ln a \left(\partial_t^2 f_j - \frac{\Delta f_j}{a^2} \right) - H\partial_t f_j 
	\right] \nonumber\\
	&&+ \Omega_{pl}^2 f_j + \Gamma \dot f_j = 0 .
	\label{df-tot}
	\ee 
	It is assumed that $f_t = 0$ and  we used 
	eqs.~(\ref{Q-HE1},\,\ref{d-F-HE-3},\,\ref{almost-fin},\,\ref{anom-eq},\,\ref{omega-rel},\,\ref{Gamma}).

	\section{Defining  $g$ and $\gamma$ system of differential equations (SoDE)}
	\label{WholeSDE}
	In total, we have ten equations for the components of  tensor $h_{\mu\nu}$ and 
	three equations for the components of vector $f_j$. In fact, only six equations for gravitational waves are linearly 
	independent.

	{In the case considered here we assume that the vector modes do not arise. The first vector $G_i$
in eqs. (\ref{Decom2}-\ref{Decom3} vanishes  due to the gauge condition $h_{0i}=0$.
The second vector $C_i$ is not zero
because the corrections to EoM contain spacial derivative of 
electromagnetic potential $\partial_{\mu}f_{\nu}$. However the vector modes decay as $a^{-2}$ and thus they 
do not play an essential role in cosmology. It worth adding that an account of one more polarisation state would
lead to a considerable complication of the system of equations. 
So in this work we confine ourselves only to scalar and tensor modes.} 
	
{Finally let us mention that the solution for tensor $h^i_j$ does not contain pure tensor mode, 
but a mixture of tensor and scalar modes. Nevertheless
the solution represents them qualitatively correctly, including the behaviour of the tensor mode that we are interested~in.}
	
{Assuming an absence of vector mode we obtain two components less in the EoM}. 
More specifically, we have two scalars, $\Phi$ and $\Psi$ (note that in the 
deal fluid model, i.e. without taking into account dissipation, $\Phi=\Psi$) and two polarisations of the tensor 
wave, that in total gives four independent equations for metric perturbations. }

{In the subsequent subsections (\ref{Simp1}, \ref{Simp2}) the SoDE is simplified for the specific choice of the reference frame, where an external magnetic field is directed along the $x$ axis.
Next, one} of two independent subsystems is solved numerically in Sec. \ref{TwoCases}.

	\subsection{{Simplification of SoDE for metric perturbations.}\label{Simp1}}
	
	{To derive the  system of equations for the metric perturbations, that is solved below, it remains to simplify the right-hand 
	side of equations (\ref{GWsys1}). 
	Let us rewrite  equation (\ref{EMTMax}-\ref{EMTHE}) for an external magnetic field is directed along the $x$ axis. 
	For individual expressions we get:}
	\begin{gather}
	F^2=2B^2,\\
	Ff=F^{\alpha}_{.\,\beta}f^{.\,\beta}_{\alpha}=F^{.\,z}_y f^y_{.\,z}+F^{.\,y}_z f^z_{.\,y}=B\left(f^y_{.\,z}-f^z_{.\,y}\right)=2Bf^y_{.\,z}\label{Ff},\\
	FFh=h^{\alpha}_{\sigma}F_{\alpha\beta}F^{\sigma\beta}=B^2\left(h^y_y+h^z_z\right)\label{FFh},\\
	\left(f^{\mu\alpha}F_{\nu\alpha}+F^{\mu\alpha}f_{\nu\alpha}\right)=\left(f^{\mu}_{.\,\alpha}F_{\nu}^{.\,\alpha}+F^{\mu}_{.\,\alpha}f_{\nu}^{.\,\alpha}\right)=B\left(f^{\mu}_{.\,z}\delta_{\nu y}-f^{\mu}_{.\, y}\delta_{\nu z}\right)+B\left(f_{\nu}^{.\,z}\delta^{\mu y}-f_{\nu}^{.\, y}\delta^{\mu z}\right),\\
	F^{\mu \alpha}F_{\nu\alpha}=2B^2\left(\delta^{\mu z}\delta_{\nu z}+\delta^{\mu y}\delta_{\nu y}\right),\\
	h^{\mu\sigma}F^{.\,\alpha}_{\sigma}F_{\nu\alpha}=B^2\left(h^{\mu}_y \delta_{\nu y}+h^{\mu}_z \delta_{\nu z}\right),\\
	h^{\alpha\lambda}F^{\mu}_{.\,\alpha}F_{\nu\lambda}=B^2\left[-h^y_z\left(\delta^{\mu z}\delta_{\nu y}+\delta^{\mu y}\delta_{\nu z}\right)+h^y_y \delta^{\mu z}\delta_{\nu z}+h^z_z \delta^{\mu y}\delta_{\nu y}\right]\label{hFF},\\
	\left(f^{\alpha\beta}F_{\lambda\alpha}F^{\sigma\lambda}F_{\sigma\beta}-h^{\alpha}_{\xi}F^{\xi\beta}F_{\lambda\alpha}F^{\sigma\lambda}F_{\sigma\beta}\right)=2B^3f^z_{.\,y}+B^4\left(h^z_z+h^y_y\right).
	\end{gather}
	In this section and in subsequent ones, for the sake of brevity, we will omit the signature $EM$ in the correction to the EMT.
	
	Using eq.(\ref{EMTMax}-\ref{EMTHE}) and eq.(\ref{Ff}-\ref{hFF}) we obtain the EMT components for $00$ (remember that $F_{0\alpha}= 0, F^y_{.\,\,z}\neq 0)$:
	\be\label{EMTBx00}
	T^{(1)}_{00}=T^{Max(1)}_{00}+T^{HE(1)}_{00}= \frac{1}{2}\left(Ff-FFh\right)-C\left[F^4 h_{00}+4F^2\left(Ff-FFh\right)\right]=\nonumber\\
	=\frac{1}{2}\left(1-16CB^2\right)\left[2Bf^y_{.\,\, z}-B^2\left(h^y_y+h^z_z\right)\right]-4CB^4 h_{00}.
	\ee
	Similarly, we derive the expression for the component $0j$:
	\be\label{EMTBx0j}
	T^{(1)}_{0j}=\left(1-8CF^2\right)f_{0\alpha}F^{\alpha}_{.\,\, j}=\left(1-16CB^2\right)B\left[f_{0y}\delta_{jz}-f_{0z}\delta_{jy}\right].
	\ee
	The expression for the $ij$ components is more cumbersome:
	\be
	&T^{i(Max1)}_j=\frac{1}{2}\delta^i_j{\left[Ff-FFh\right]}+h^i_{\sigma}F^{\sigma\alpha}F_{j\alpha}+{h^{\sigma}_{\lambda}F^{i\lambda}F_{j\sigma}}+{\left(F^i_{.\,\,\lambda}f^{\lambda}_{.\,\, j}+f^i_{.\,\,\lambda}F^{\lambda}_{.\,\, j}\right)},\\
	&T^{i(HE 1)}_j=-h^i_j4CB^4-8CB^2\delta^i_j{\left(Ff-FFh\right)}-16B^2C{\left(F^i_{.\,\,\lambda}f^{\lambda}_{.\,\, j}+f^i_{.\,\,\lambda}F^{\lambda}_{.\,\, j}\right)}\nonumber\\
	&+16CF^{i\lambda}F_{j\lambda}{\left(Ff-FFh\right)}-16B^2C{h^{\sigma}_{\lambda}F^{i\lambda}F_{j\sigma} }
	\ee
	Let's {regroup} the terms
	\be
	&T^{i(1)}_j=\left[\frac{1}{2}\left(1-16CB^2\right)\delta^i_j+16CF^{i\lambda}F_{j\lambda}\right]\left(Ff-FFh\right)\nonumber\\
	&+\left(1-16CB^2\right)\left[F^i_{.\,\,\lambda}f^{\lambda}_{.\,\, j}+f^i_{.\,\,\lambda}F^{\lambda}_{.\,\, j}+h^{\sigma}_{\lambda}F^{i\lambda}F_{j\sigma} \right]\nonumber\\
	&-4CB^4h^i_j+h^i_{\sigma}F^{\sigma\alpha}F_{j\alpha}=\nonumber\\
	&=\left[\frac{1}{2}\left(1-16CB^2\right)\delta^i_j+16CB^2\left(\delta^{iy}\delta_{jy}+\delta^{iz}\delta_{jz}\right)\right]\left[2Bf^y_{.\,\,z}-B^2\left(h^y_y+h^z_z\right)\right]\nonumber\\
	&+\left(1-16CB^2\right)B\left[f^z_{.\,\,j}\delta^{iy}-f^y_{.\,\,j}\delta^{iz}-f^i_{.\,\,z}\delta_{jy}+f^i_{.\,\,y}\delta_{jz}\right]\nonumber\\
	&+\left(1-16CB^2\right)B^2\left[-h^y_z\left(\delta^{i z}\delta_{j y}+\delta^{i y}\delta_{j z}\right)+h^y_y \delta^{i z}\delta_{j z}+h^z_z \delta^{i y}\delta_{j y}\right]\label{hFF}\nonumber\\
	&-4CB^4h^i_j+B^2\left(h^i_y \delta_{jy}+h^{i}_{z}\delta_{jz}\right).
	\ee
	It will now be useful to write down the spatial components separately. After reducing similar terms we get
	\be\label{EMTBx1}
	&T^{x(1)}_x=\frac{1-16CB^2}{2}\left[2Bf^y_{.\,\,z}-B^2\left(h^y_y+h^z_z\right)\right]-4CB^4h^x_x,\\
	&T^{x(1)}_y=-B\left(1-16CB^2\right)f^x_{.\,\,z}+B^2\left(1-4CB^2\right)h^x_y,\\
	&T^{x(1)}_z=B\left(1-16CB^2\right)f^x_{.\,\,y}+B^2\left(1-4CB^2\right)h^x_z,\\
	&T^{y(1)}_y=B\left(-1+48CB^2\right)f^y_{.\,\,z}+B^2\frac{h^y_y+h^z_z}{2}-12CB^4\left(h^y_y+2h^z_z\right),\\
	&T^{y(1)}_z=12CB^4h^y_z,\\
	&T^{z(1)}_z=B\left(-1+48CB^2\right)f^y_{.\,\,z}+B^2\frac{h^y_y+h^z_z}{2}-12CB^4\left(2h^y_y+h^z_z\right).\label{EMTBx6}
	\ee
	System of equations (\ref{GWsys1}) is now  rewritten as equations(\ref{EMTBx00},\ref{EMTBx0j}, \ref{EMTBx1}-\ref{EMTBx6}).
	
	\subsection{{Simplification of SoDE for electromagnetic waves }\label{Simp2}}
	
	Let us simplify the system of equations for an electromagnetic wave for the case when the external 
	magnetic field is directed along the $x$ axis.
	For spatial components equation (\ref{df-tot}) was derived  in general form.
	We assume that $C(T)=const$, so the time derivative is $\dot{C}=0$. Also, to begin with, let’s omit the last three terms 
	{in} the left side of the equation, {that} 
	take into account the interaction of photons with plasma. Now let’s write the convolutions with the 
	background tensor of the electromagnetic field in terms of field $\boldsymbol{B}$:
	\be
	&-  g_{\nu j}   \left(h^{\mu \alpha}  \partial_\mu  F^\nu_{.\alpha} + 	F^\mu_{. \alpha} \partial_\mu h^{\nu\alpha}\right)=g_{\nu j}\left[h^{0y}\delta^{\nu z}-h^{0z}\delta^{\nu y}\right]\dot{B}+g_{\nu j}B\left[\partial_z h^{\nu y}-\partial_y h^{\nu z}\right]=\nonumber\\
	&=\left[h^{0}_{y}\delta_{j z}-h^{0}_{z}\delta_{j y}\right]\dot{B}+B\left[\partial_z h^{y}_j-\partial_y h^{z}_j\right],\\
	&\left(f^{\mu}_{. j} -h^{\mu\alpha}F_{\alpha j}-h_{j \alpha}F^{\mu\alpha}\right)\partial_{\mu}B^2=\left(f^{t}_{. j} -h^{0}_{y} \delta_{jz}B+h^0_z \delta_{jy} B
	\right)2B\dot{B},\\
	&F^{\mu}_{. j} \partial_{\mu}\left(fF-FFh\right)=B^2\left[\delta_{jz}\partial_y-\delta_{jy}\partial_z\right]\left[2f^y_{.\,\,z}-B\left(h^y_y+h^z_z\right)\right],\\
	&\left[(\dot{C}+HC)\delta^t_{\mu}+C\partial_{\mu}\right]\tilde{F}^{\mu}_{. j}(f\tilde{F}) =C\partial_m\tilde{F}^m_{.\,\,j}(f\tilde{F}) =0.
	\ee
	The last equality follows from the fact that the only non-zero component of the dual electromagnetic tensor in 
	the case when the magnetic field is directed along the $x$ {is}  $\tilde{F}_{0x}=-{aB}/{2 }\neq0$.

	So, let’s write the resulting equation (we omit the terms from the interaction with the plasma):
	\be
	 &&\left( 1 -  16 C B^2\right) \left[ \partial_t^2 f_j - \frac{\Delta f_j}{a^2} + \frac{\partial_j \partial_k f_k}{a^2}
	+ H \partial_t f_j +\left[h^{0}_{y}\delta_{j z}-h^{0}_{z}\delta_{j y}\right]\dot{B}+B\left[\partial_z h^{y}_j-\partial_y h^{z}_j\right]  \right]  \nonumber \\
	 &&- 16 C \left[ \left(f^{t}_{. j} -h^{0}_{y} \delta_{jz}B+h^0_z \delta_{jy} B
	\right)2B\dot{B} \right.
	 \nonumber \\
	&&+ \left. B^2\left[\delta_{jz}\partial_y-\delta_{jy}\partial_z\right]\left[2f^y_{.\,\,z}-B\left(h^y_y+h^z_z\right)\right]\right]  = 0.
	\label{d2-f}
	\ee	
	By analogy with the equations for gravitational waves, we write for $x$, $y$, and $z$ components, respectively:
	\be
	&& \left( 1 -  16 C B^2\right) \left[ \partial_t^2 f_x - \frac{\Delta f_x}{a^2} + \frac{\partial_x \partial_k f_k}{a^2}
	+ H \partial_t f_x +B\left[\partial_z h^{y}_x-\partial_y h^{z}_x\right]  \right]
	\nonumber \\
	&& - 32CB\dot{B}f^t_{.\,\,x} = 0,
	\ee
	\be
	& \left( 1 -  16 C B^2\right) \left[ \partial_t^2 f_y - \frac{\Delta f_y}{a^2} + \frac{\partial_y \partial_k f_k}{a^2}
	+ H \partial_t f_y -h^{0}_{z}\dot{B}+B\left[\partial_z h^{y}_y-\partial_y h^{z}_y\right]  \right]  \nonumber \\
	& - 16 C \left[ \left(f^{t}_{. y} +h^0_z B
	\right)2B\dot{B}
	-B^2\partial_z\left[2f^y_{.\,\,z}-B\left(h^y_y+h^z_z\right)\right]\right]  = 0,
	\ee
	\be
	& \left( 1 -  16 C B^2\right) \left[ \partial_t^2 f_z - \frac{\Delta f_z}{a^2} + \frac{\partial_z \partial_k f_k}{a^2}
	+ H \partial_t f_z +h^{0}_{y}\dot{B}+B\left[\partial_z h^{y}_z-\partial_y h^{z}_z\right]  \right]  \nonumber \\
	& - 16 C \left[ \left(f^{t}_{. z} -h^{0}_{y} B
	\right)2B\dot{B}
	+ B^2\partial_y\left[2f^y_{.\,\,z}-B\left(h^y_y+h^z_z\right)\right]\right]  = 0.
	\ee
	Similar ones can be given,  taking into account that $B\sim {1}/{a^2} $ 
	(meaning $\dot{B}=-2HB$) and that $f_t=0$. For $x$ components we get:
	\be\label{emwX}
	& \left( 1 -  16 C B^2\right) \left[ \ddot{f_x} - \frac{\Delta f_x}{a^2} + \frac{\partial_x \partial_k f_k}{a^2}
	+ H \dot{f_x} +B\left[\partial_z h^{y}_x-\partial_y h^{z}_x\right]  \right] +64CB^2H\dot{f_{x}} = 0,
	\ee
	or
	\be
	& \left( 1 -  16 C B^2\right) \left[ \ddot{f_x} - \frac{\Delta f_x}{a^2} + \frac{\partial_x \partial_k f_k}{a^2}
	+B\left[\partial_z h^{y}_x-\partial_y h^{z}_x\right]  \right] +\left(1+48CB^2\right)H\dot{f_{x}} = 0.
	\ee
	For $y$ component:
	\be\label{emwY}
	& \left( 1 -  16 C B^2\right) \left[ \ddot{f_y} - \frac{\Delta f_y}{a^2} + \frac{\partial_y \partial_k f_k}{a^2}
	+ H \dot{f_y} +2HBh^{0}_{z}+B\left[\partial_z h^{y}_y-\partial_y h^{z}_y\right]  \right]  \nonumber \\
	& + 16 CB^2 \left[4H \left(\dot{f_{y}} +h^0_z B
	\right)
	+\partial_z\left[2f^y_{.\,\,z}-B\left(h^y_y+h^z_z\right)\right]\right]  = 0,
	\ee
	or
	\be
	& \left( 1 -  16 C B^2\right) \left[ \ddot{f_y} - \frac{\Delta f_y}{a^2} + \frac{\partial_y \partial_k f_k}{a^2}
	 \right]+\left(1+48CB^2\right)H\dot{f_y}-32CB^2\frac{\partial_zf_{yz}}{a^2}\nonumber\\
	 &-B\left(1-16CB^2\right)\partial_y h^z_y-16CB^3\partial_z h^z_z+B\left(1-32CB^2\right)\partial_z h^y_y = 0.
	\ee
	For $z$ component:
	\be\label{emwZ}
	& \left( 1 -  16 C B^2\right) \left[ \ddot{ f_z} - \frac{\Delta f_z}{a^2} + \frac{\partial_z \partial_k f_k}{a^2}
	+ H \dot{ f_z} -2HBh^{0}_{y}+B\left[\partial_z h^{y}_z-\partial_y h^{z}_z\right]  \right]  \nonumber \\
	& +16 C B^2\left[ 4H\left(\dot{f_{z}} -h^{0}_{y} B
	\right)
	- \partial_y\left[2f^y_{.\,\,z}-B\left(h^y_y+h^z_z\right)\right]\right]  = 0,
	\ee
	or
	\be
	& \left( 1 -  16 C B^2\right) \left[ \ddot{ f_z} - \frac{\Delta f_z}{a^2} + \frac{\partial_z \partial_k f_k}{a^2}
	 \right] +\left(1+48CB^2\right)H\dot{f_z}+32CB^2\frac{\partial_yf_{yz}}{a^2} \nonumber\\
	&+B\left(1-16CB^2\right)\partial_zh^y_z+16CB^3\partial_y h^y_y-B\left(1-32CB^2\right)\partial_y h^z_z= 0.
	\ee
	
	Next we would like to show the validity of the requirement $f_t=0$. In general, due to the homogeneity of the 
	magnetic field (depending only on time), we arrive at the following equation for the time component:
	\be
	\left(1-16CB^2\right)\partial_{\mu}f^{\mu 0}=0.
	\ee
	Now we need to select a calibration. If our problem can be called magnetostatic, in such cases the Coulomb 
	gauge $\partial_{\mu}f^{\mu}=0$ is usually introduced, then we get:
	\be
	&\partial_{\mu}f^{\mu 0}=\partial_{\mu}\partial^{\mu}f_0+\partial_0\partial_{\mu}f^{\mu}=\partial_{\mu}\partial^{\mu}f_0,\nonumber\\
	&\partial_{\mu}\partial^{\mu}f_0=0=>f_0=const.
	\ee
	From the initial conditions of electrical neutrality we find that this constant is equal to zero.

	\section{{Two examples of  gravitational wave directions}} \label{TwoCases}
	
For any initial direction of the gravitational (tensor) wave propagation, we can decompose it into a parallel and perpendicular component relative to the external magnetic field. {Note that we consider the case when
an initial pure tensor plane wave propagates from vacuum into a region with a 
magnetic field (and, in the future, with plasma).}
	
	{It is  shown below that,} for $\mathbf{k} || \mathbf{B}$ the scalar mode of metric perturbations is not excited, 
	{and the electromagnetic wave is not excited as well.} 
	For the perpendicular component $\mathbf{k} \bot \mathbf{B}$ the situation is different - the scalar mode of metric 
	perturbations and both polarizations of the electromagnetic wave are excited. Until now, we have not taken into account 
	dissipation and loss of coherence for photons due to their interaction with plasma. But even without taking these phenomena 
	into account, it is already possible to detect a change in the amplitude of the initial tensor GW due to the transition to the 
	scalar mode of metric disturbances and to an electromagnetic wave. We will consider both of these cases in more detail 
	in the next two subsections.
	
	\subsection{$\mathbf{k} || \mathbf{B}$}
	Let us write down the basic relations that {allow to}
	 simplify the system of differential equations for metric perturbations and for an electromagnetic wave (EMW):
	\be
	&\mathbf{k}=\left(k_x,0,0\right),\\
	&f^{\mu}=\left(0,0,f^y,f^z\right),\\
	&f^{\mu}(t,x) \sim \exp{(ik_x x)},\\
	&h_{\mu\nu} (t,x)\sim \exp{(ik_x x)},
	\ee
	
	\begin{gather}
	h^{\mu}_{\nu}=\begin{bmatrix}
	0 & 0 & 0 & 0\\
	0 & 0 & 0 & 0\\
	0 & 0 & h_+ & h_{\times}\\
	0 & 0 & h_{\times} & -h_+
	\end{bmatrix},
	\end{gather}
	
	\be
	h^y_y+h^z_z=0.
	\ee
	 Taking into account what was written above, we write the system of equations in terms of $h_+,h_{\times}$. From equation~(\ref{EMTBx00}, \ref{EMTBx0j}, \ref{EMTBx1}-\ref{EMTBx6}) we obtain
	 \be
	 &T^{(1)}_{00}=0,\\
	 &T^{(1)}_{0j}=\left(1-16CB^2\right)B\left[\dot{f_y}\delta_{jz}-\dot{f_z}\delta_{jy}\right],\\
	 &T^{x(1)}_x=0,\\
	 &T^{x(1)}_y=-B\left(1-16CB^2\right)\partial^x f_z,\\
	 &T^{x(1)}_z=B\left(1-16CB^2\right)\partial^x f_y,\\ \label{parallelEmtYY}
	 &T^{y(1)}_y=12CB^4h_+\\ \label{parallelEmtYZ}
	 &T^{y(1)}_z=12CB^4h_{\times},\\ \label{parallelEmtZZ}
	 &T^{z(1)}_z=-12CB^4 h_{+}.
	 \ee
	 From equations~(\ref{emwX}, \ref{emwY}, \ref{emwZ}) we obtain, respectively
	 \be\label{parallelEmwX}
	& \left(1-16CB^2\right)\left[\ddot{f_x}-\frac{\Delta f_x}{a^2}+\frac{\partial_x^2 f_x}{a^2}+H\dot{f_x}\right]+64CB^2H\dot{f}_x=0,\\ \label{parallelEmwY}
	 &\left(1-16CB^2\right)\left[\ddot{f_y}-\frac{\Delta f_y}{a^2}+H\dot{f_y}\right]+64CB^2H\dot{f}_y=0,\\ \label{parallelEmwZ}
	 &\left(1-16CB^2\right)\left[\ddot{f_z}-\frac{\Delta f_z}{a^2}+H\dot{f_z}\right]+64CB^2H\dot{f}_z=0.
	 \ee
	 In equation (\ref{parallelEmwX}-\ref{parallelEmwZ}) there are no terms related to the gravitational wave, therefore, if the electromagnetic wave was not initially present, it does not arise for the case when the wave vector is parallel to the external magnetic field. {Hence we obtain that}
	 \be
	 &T^{(1)}_{0j}=0,\\
	 &T^{x(1)}_y=0,\\
	 &T^{x(1)}_z=0.
	 \ee
	 From the remaining non-zero components of EMT eq. (\ref{parallelEmtYY}-\ref{parallelEmtZZ}) we see that the GW configuration is preserved: it remains tensorial and no scalar modes arise.
	 
	\subsection{$\mathbf{k} \bot \mathbf{B}$}
	
	Similar to the previous {subsection, we} write down the main relations that will help {to} simplify the system of 
	differential equations for perturbations of metric and for electromagnetic waves. Let's direct the wave vector along the 
	$z$ axis (it is always possible to rotate the coordinate system so that $k_y=0$).
	\be
	&\mathbf{k}=\left(0,0,k_z\right),\\
	&f^{\mu}=\left(0,f^x,f^y,0\right),\\
	&f^{\mu}(t,x) \sim \exp{(ik_z z)},\\
	&h_{\mu\nu} (t,x)\sim \exp{(ik_z z},
	\ee
	
	\begin{gather}
	h^{\mu}_{\nu}=\begin{bmatrix}
	0 & 0 & 0 & 0\\
	0 & h_+ & h_{\times} & 0\\
	0 & h_{\times} & -h_+ & 0\\
	0 & 0 & 0 & 0
	\end{bmatrix},
	\end{gather}
	
	\be
	h^y_y+h^z_z=-{h_{+}}.
	\ee
	 Taking into account what was written above, we will write the system of equations in terms of $h_+,h_{\times}$. From equation~(\ref{EMTBx00}, \ref{EMTBx0j}, \ref{EMTBx1}-\ref{EMTBx6}) we obtain
	 \be\label{perpendEmt00}
	 &T^{(1)}_{00}=\frac{1}{2}\left(1-16CB^2\right)\left[-2B\partial_z f^y+B^2 h_{+}\right],\\\label{perpendEmt0j}
	 &T^{(1)}_{0j}=\left(1-16CB^2\right)B\dot{f}_y\delta_{jz},\\\label{perpendEmtXX}
	 &T^{x(1)}_x=\frac{1}{2}\left(1-16CB^2\right)\left[-2B\partial_z f^y+B^2 h_{+}\right] -4CB^4h_+,\\\label{perpendEmtXY}
	 &T^{x(1)}_y=B\left(1-16CB^2\right)\partial_z f^x +B^2\left(1-4CB^2\right)h_{\times},\\\label{perpendEmtXZ}
	 &T^{x(1)}_z=0,\\\label{perpendEmtYY}
	 &T^{y(1)}_y=B\left(1-48CB^2\right)\partial_z f^y-\frac{B^2}{2}h_+ + 12CB^4h_+\\\label{perpendEmtYZ}
	 &T^{y(1)}_z=0,\\ \label{perpendEmtZZ}
	 &T^{z(1)}_z=B\left(1-48CB^2\right)\partial_z f^y-\frac{B^2}{2}h_++24CB^4 {h_{+}}.
	 \ee
	 
	 From Eq.~(\ref{emwX}, \ref{emwY}, \ref{emwZ}) respectively
	 \be\label{perpendEmwX}
	& \left(1-16CB^2\right)\left[\ddot{f_x}-\frac{\Delta f_x}{a^2}+B\partial_z h_{\times}\right]+\left(1+48CB^2\right)H\dot{f}_x=0,\\ \label{perpendEmwY}
	 &\left(1-16CB^2\right)\left[\ddot{f_y}-\frac{\Delta f_y}{a^2}+H\dot{f_y}-B\partial_z h_{+}\right]+\nonumber\\ 
	 &+16CB^2\left[4H\dot{f}_y+\partial_z\left(-2\partial_zf^y+Bh_{+}\right)\right]=0,\\ \label{perpendEmwZ}
	 &\left(1-16CB^2\right)\left[\ddot{f_z}-\frac{\Delta f_z}{a^2}+H\dot{f_z}\right]+64CB^2H\dot{f}_z=0.
	 \ee
	In equation (\ref{perpendEmwZ}) there are no terms associated with a gravitational wave. So, as 
	expected, longitudinal EMW does not arise. From {equations} (\ref{perpendEmwX},\ref{perpendEmwY}) 
	it follows that an electromagnetic wave with polarization along the $x$ axis is generated by the polarization $h_{\times}$ of the GW, and an electromagnetic wave with polarization along the $y$ axis - polarization $h_+$ of the gravitational wave. Also from the equations (\ref{perpendEmt00}-\ref{perpendEmtZZ}) we clearly see the emergence of a scalar mode from the equations for the $00$ and $zz$ components of the EMT (see equation (\ref{GWsys1})).
	 
	 It is important to note that 
	 the expressions for EMT in terms of $h_+,h_{\times}$ are valid only at the moment of time immediately following the 
	 initial moment of GW entry into the region with a magnetic field. Further, the wave ceases to be purely tensorial, and it is 
	 impossible to assert that $h^x_x=-h^y_y$. To find a solution, it is necessary to express  
	 all the quantities 
	 precisely in terms of $h^x_x$ 
	 and $h^y_y$ (not in terms of $h_+$) or in terms of expansion in helicity, introducing $\Phi$ and $\Psi$.
	 
	 \section{System solution in the case $\mathbf{k} \bot \mathbf{B}$} \label{Sol}
	 
	 Let us write out the system of equations completely, taking into account the conclusions of the previous section that $f^z,\, h_x^z,\, h_y^z$ components that are absent at the beginning do not arise during the conversion of tensor GW into photons and scalar perturbations of the metric.
	 
	 Let us draw the reader's attention to the fact that we write the equations in terms of the electromagnetic potential with the superscript $f^{\mu}$ and the gravitational wave potential with mixed indices $h_{\mu}^{\nu}$. In this case, we use the following expansion in helicity {states} for perturbation of the metric
	 \be
	&h_{tt}=2\Phi(t,\bf{r}),\\
	&h^z_{z} = 2 \Psi(t,\bf{r}),\\
	&h^x_x = 2 \Psi(t,\bf{r})+ h_{+}(t,\bf{r}),\\
	&h^y_y = 2 \Psi(t,\bf{r})- h_+(t,\bf{r}),\\
	&h^x_y= h_{\times}(t,\bf{r}).
	 \ee
	 {We also make the Fourier expansion in terms of momentum, and accept the law of the scale factor 
	 variation} with time, 
	 {corresponding to the stage of radiation dominance $a(t) \sim t^{\frac{1}{2}}$.}

		 To further search for a numerical solution, it would be {convenient}
		 to introduce dimensionless quantities. To do this, let us change the notation
	 \be
	 f^x/m_{pl}\rightarrow  f^x,\\
	 f^y/m_{pl}\rightarrow f^y,
	 \ee
	 and introduce $\tau_0$  to make the scale factor {dimensionless}
	 \be
	 a=\sqrt{\frac{t}{\tau_0}}.
	 \ee
	 Due to the last change, the tensor $h_{\mu\nu}$ also becomes dimensionless.
	 
	{Let us assume that \tcviol{at} the present-day Universe $a_0=1$. 
	 {This is just a choice of reference point and this choice does not influence the solution,} 
	 because the constant factor in front of the scale factor function has no physical meaning. 
	The condition is convenient in our problem to recalculate magnetic field strength using the present day magnitude.
	Using the scale factor dependence during matter dominance epoch 
	{$a=\left({t}/{\tau_{tot}}\right)^{2/3}$} we 
	obtain for the coefficient $\tau_0$: 
	\be
	\tau_0=\tau_{tot}\left(\frac{\tau_{tot}}{t_{eq}}\right)^{1/3}\approx 35\,\,  \tau_{tot},
	\ee
	{where $\tau_{tot}=13.8 \times10^9$ years is the age of the Universe, 
	$t_{eq}=3.3 \times10^5$  years is the moment when radiation and dust energy densities were equal.}}

	 We accept also
	 that the scale factor $a$ varies in the interval $10^{-9} \leqq a \leqq 10^{-4}$. The selected interval lies inside
	  the radiation dominance epoch (from the hadronic to the recombination).
	 
	 For the magnitude of the magnetic field $B_0$ in the system of equations, we take its value at the present time. 
	 There are {bounds obtained from observations:}
	 	 $10^{-16} \leqq B_0 \leqq 10^{-9}$ Gs \cite{MagFieldLimitations}. 
	{ Therefore, let's put $B_0=10^{-9} \times1.95 \times10^{-14}\,\, \text{MeV}^2$.}
	 
	 After the Fourier transform {over}
	 momentum, it will be clear that the system of equations contains both imaginary and real terms, therefore, to solve the SDE numerically, it will be necessary to decompose each of the required quantities into real and imaginary parts. For example $h_+=Re(h_+)+i Im(h_+)$ and so on.
	 For brevity, we write down systems of equations without dividing into real and imaginary parts.
	 To obtain a more universal result, it is convenient to write the system in terms of $a(t)$. The first independent system has the following form:
	 
	  \begin{align}
	  \label{SDEtoSolve1}
	 &f^x: a^2H^2{f}^{x \prime\prime}+aH^2\left[1+a\frac{H^{\prime}}{H}+8\frac{2B_0^2C_0-a^4}{16B_0^2C_0-a^4}+aH\Gamma\right]f^{x \prime}+\nonumber\\
	 &+\left[\frac{k^2}{a^2}+2aHH^{\prime}-8H^2\frac{4B_0^2C_0+a^4}{16B_0^2C_0-a^4}+2\Gamma H+\omega_{pl}^2\right]f^x{-\alpha\beta H^2\left(af^{x\prime}+2f^x\right)}\nonumber\\
	 &{+\alpha\beta \ln{a}\left[a^2H^2f^{x\prime\prime}+\left(\frac{k^2}{a^2}+2aHH^{\prime}\right)f^x+\left(5aH^2+a^2HH^{\prime}\right)f^{x\prime}\right]}=-\frac{ikB_0}{a^4 m_{pl}}h_{\times},\nonumber\\
	 &h_x^y: a^2 H^2 {h}_{\times}^{\prime\prime}+\left(4aH^2+a^2HH^{\prime}\right){h}_{\times}^{\prime}+\left[\frac{k^2}{a^2}+\frac{16\pi G B_0^2}{{ a^4}}\left(1-\frac{4B_0^2 C_0}{a^4}\right)\right]h_{\times} =\nonumber\\
	 &=- \frac{16\pi GB_0 ik}{{a^2}}\left(1-\frac{16B_0^2C_0}{a^4}\right)m_{pl}f^x,
	 \end{align}
	 where the prime denotes the derivative with respect to the scale factor,
	 and we introduced the attenuation of the electromagnetic wave due to its interaction with the plasma using 
	 the damping factor $\Gamma \propto \alpha^2 T(a)$ and the plasma frequency $\omega_{pl}^2 \propto \alpha T(a)$.  Let us recall that $T(a)\propto  {1}/{a}$.

	 For all solution interval terms with the multiplier $\alpha\beta \ln{a}$ can be neglected. Therefore the resulting system is
	  \begin{align}
	  \label{SDEtoSolve}
	 &f^x: a^2H^2{f}^{x \prime\prime}+aH^2\left[1+a\frac{H^{\prime}}{H}+8\frac{2B_0^2C_0-a^4}{16B_0^2C_0-a^4}+aH\Gamma\right]f^{x \prime}+\nonumber\\
	 &+\left[\frac{k^2}{a^2}+2aHH^{\prime}-8H^2\frac{4B_0^2C_0+a^4}{16B_0^2C_0-a^4}+2\Gamma H+\omega_{pl}^2\right]f^x{-\alpha\beta H^2\left(af^{x\prime}+2f^x\right)}=-\frac{ikB_0}{a^4 m_{pl}}h_{\times},\nonumber\\
	 &h_x^y: a^2 H^2 {h}_{\times}^{\prime\prime}+\left(4aH^2+a^2HH^{\prime}\right){h}_{\times}^{\prime}+\left[\frac{k^2}{a^2}+ \frac{16\pi G B_0^2}{{a^4}}\left(1-\frac{4B_0^2 C_0}{a^4}\right)\right]h_{\times} =\nonumber\\
	 & = - \frac{16\pi GB_0 ik}{{a^2}}\left(1-\frac{16B_0^2C_0}{a^4}\right)m_{pl}f^x,
	 \end{align}

	 It can be seen that the initial conditions
	 \be
	 \label{InitCond}
	 h_{\times}(0)&=&h^0_{\times},\nonumber\\
	 f^y(0)&=&0,\nonumber\\
	 h_{\times}^{\prime}(0)&=&{0},\nonumber\\
	 f^{y\prime}(0)&=&0
	 \ee
	 give nontrivial solution. 
	 
	 {Let us stress here that the chosen initial conditions just allow us to formulate a simple problem to solve and to obtain the effect in order of magnitude, i.e. to obtain a representative result. That is the first step of the investigation. In future works we are going to take step by step more close to the real physics conditions.}
	 
	 {It is important to note that there are poles at $a^4=16B_0^2C_0$ in the first equation. Let us remind that the effective Heisenberg-Euler action eq.(\ref{A-HE}) is correct under the assumption of a weak external electromagnetic field. In our case that means that $B_0/a^2 \ll m_e^2$. {This restirction is valid 
in the selected interval of the variation of the scale factor: 
$a \in [10^{-9},\,10^{-4}]$. Indeed, for $B_0=1$ nGs inside the interval for the scale factor we obtain
	 \be
	 B \in [10, 10^9] \, \text{Gs} = [1.95*10^{-13}, 1.95*10^{-3}]\, \text{MeV}^2.
	 \ee
	 All the values are less than the electron mass squared $m_e^2 \approx 0.25$ MeV$^2$.  Using the definition (\ref{C0definition}) we can rewrite the term 
	 proportional to  $B_0^2C_0 / a^4$ as:}
	 \be
	 \frac{C_0 B_0^2}{a^4} = \frac{\alpha^2}{90\,m_e^4}B^2.
	 \ee
	 {On the other hand,
	 we have shown that $B\ll m_e^2$  in the whole interval of solution variation}. Therefore, after squaring and multiplying by ${\alpha^2}/{90} \ll 1$, the condition 
	 remains the same
	 \be
	 \frac{\alpha^2}{90\,m_e^4}B^2 \ll 1,
	 \ee
	 and means that the  {correction to the Maxwell action proportional to $\alpha^2$  is sufficiently accurate} for our consideration.}
	 
	 {Another important question to be solved in future work is to what minimum value of the scale factor should the solution be expanded? The solution to this question 
	 should be { looked for  in the theories of cosmological magnetogenesis, 
	 that study the epoch when the cosmological magnetic field was generated. 
	 We should also} keep in mind that as the scale factor approaches the pole,
	 higher order corrections will be excited \cite{HE}, thereby removing any potential pole.}

	 The second subsystem, which involves the quantities $\{ \Phi,\,\Psi, \,f^y,\,h_+\} $, is larger, more complex and requires solving 
	 many sub-problems. For example, the question of whether scalar metric perturbations propogate 
	 is quite nontrivial and requires careful analysis. In order not to make the article too cumbersome, in this work we will concentrate on solving only the 
	 subsystem $\{f^x,\,h_{\times}\}$.
	 	 
	In order to numerically solve the system we need to divide both parts of the equation by $a^2H^2$ and to introduce two new functions to lower the order of the equation in order to make it look like: $y'=f(x,y)$.	 \par
	 	Two new functions and a system $4\times 4$ to be solved:
	\begin{align}
	&{{f}^{x \prime}}=v_{f_x},\\
	&{{h}_{\times}^{\prime}}=v_{h_{\times}},\\
	&f^x: v_{f_x}'=-\frac{1}{a}\left[1+a\frac{H^{\prime}}{H}+8\frac{2B_0^2C_0-a^4}{16B_0^2C_0-a^4}+aH\Gamma \right]v_{f_x}{+\frac{\alpha\beta}{a^2}\left(af^{x\prime}+2f^x\right)}-\nonumber\\
	&-\frac{1}{a^2H^2}\left[\frac{k^2}{a^2}+2aHH^{\prime}-8H^2\frac{4B_0^2C_0+a^4}{16B_0^2C_0-a^4} +2\Gamma H+\omega_{pl}^2\right]f^x-\frac{ikB_0}{a^6H^2 m_{pl}}h_{\times}\nonumber\\
	&h_x^y:  v_{h_{\times}}'=-\frac{\left(4aH^2+a^2HH^{\prime}\right)}{a^2 H^2}v_{h_{\times}}-\frac{1}{a^2 H^2}\left[\frac{k^2}{a^2}+\frac{16\pi G B_0^2}{{a^4}}\left(1-\frac{4B_0^2 C_0}{a^4}\right)\right]h_{\times}-\nonumber\\
	&-\frac{16\pi GB_0 ik}{{a^2}}\left(1-\frac{16B_0^2C_0}{{a^4}}\right)\frac{m_{pl}f^x}{a^2 H^2},
	\end{align}	
	We use fifth-order implicit Runge-Kutta method which is algebraically stable and allows solving stiff systems of differential equations, see for more\cite{Runge}.

	\subsection{{Method of solution validation}}
	
	Before solving the system for a non-zero magnetic field strength, we must check whether the method for the SoDE solving works correctly for the case when it is absent. Equation of motion for tensor gravitational waves in the approximation $k \eta \sim 1$, where $\eta(t)=\int \frac{dt}{a(t)}$ is a conformal time, can be solved analitically. The solution has the form
	\be
	h(\eta)=h_{init} \frac{\sin{\left(k \eta + \phi_0\right)}}{k\eta},
	\ee
	where $h_{init}$ is an initial magnitude of tensor perturbation, and $\phi_0$ is a constant phase. The last two parameters are defined from the matching with the constant mode, obtained from the EoM solution in the approximation $k\eta \ll 0$ (see Sec. 3.2 in Ref.\cite{GR-2}).
	
\begin{figure}[hbt!]
	\begin{subfigure}{\textwidth}
  	\centering
 	 \includegraphics[width=.8\linewidth]{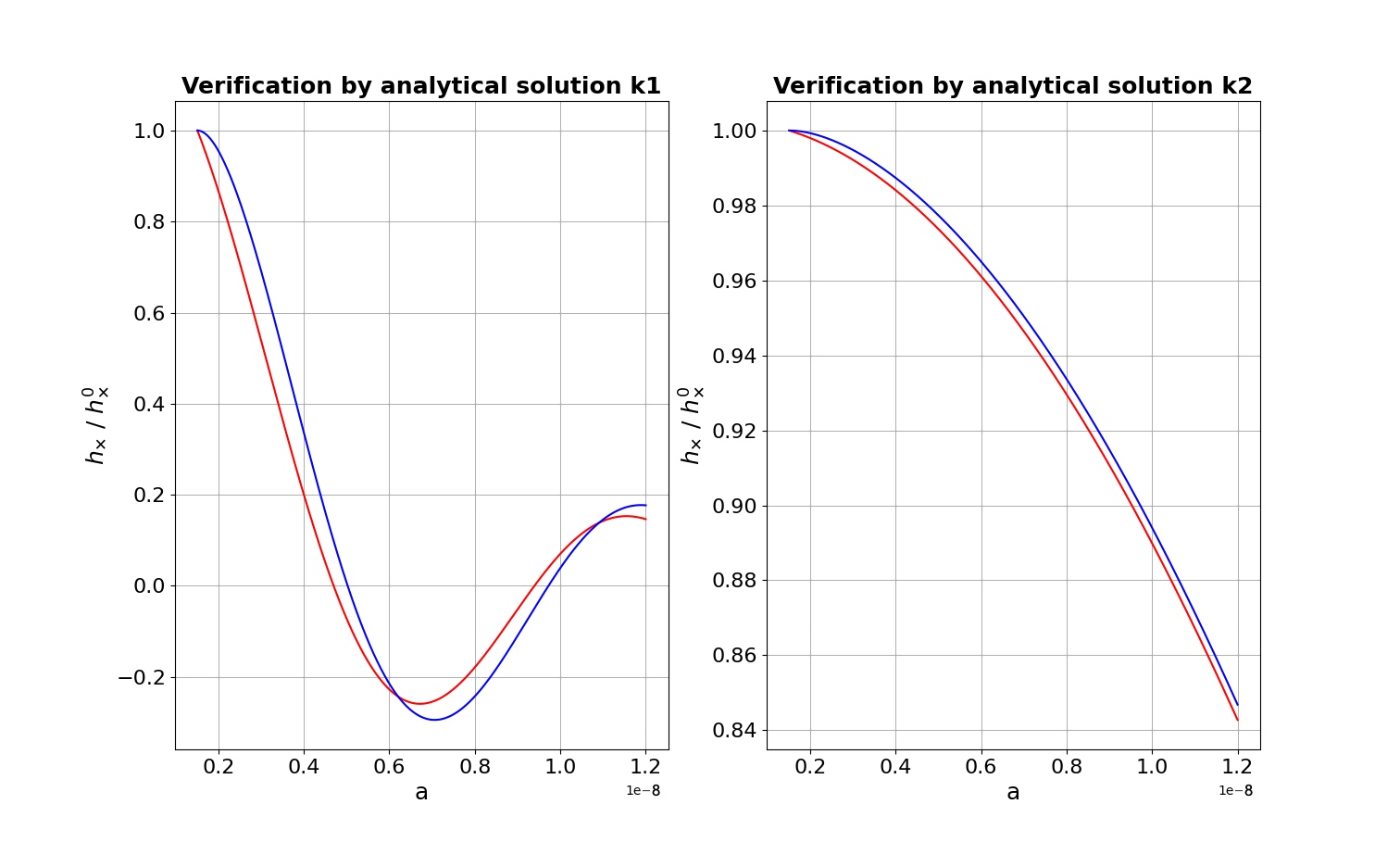}
  	\caption{without phases}
 	 \label{fig:sfig1}
	\end{subfigure}\\
	\begin{subfigure}{\textwidth}
	  \centering
  	\includegraphics[width=.8\linewidth]{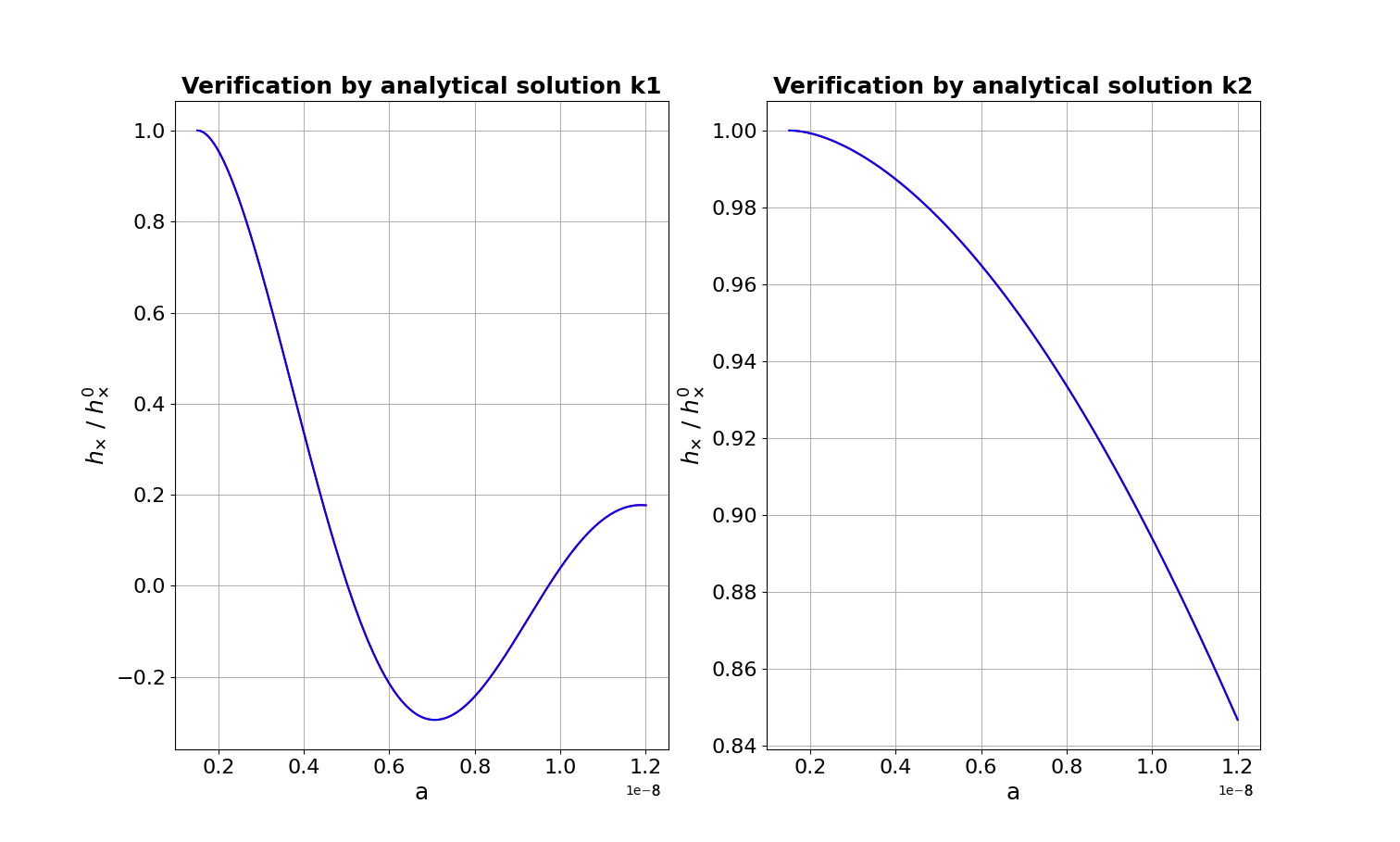}
	  \caption{with phases}
 	 \label{fig:sfig2}
	\end{subfigure}
	\caption{Verification of the numerical solution (blue line) by the analytical solution (red line) for two frequencies $k_1$ (left panel) and $k_2$ (right panel)}
	\label{fig:1}
\end{figure}

	In the Fig. \ref{fig:sfig1} the numerical and analytical solutions are presented for the two conformal time values $\eta_1 =13.2, \eta_2=105.6$ and corresponding to them frequencies $k_1=0.076, k_2=0.0095$ Hz satisfying the condition $k \eta \sim 1$.

\begin{figure}
	\centering
 		\includegraphics[scale=0.4]{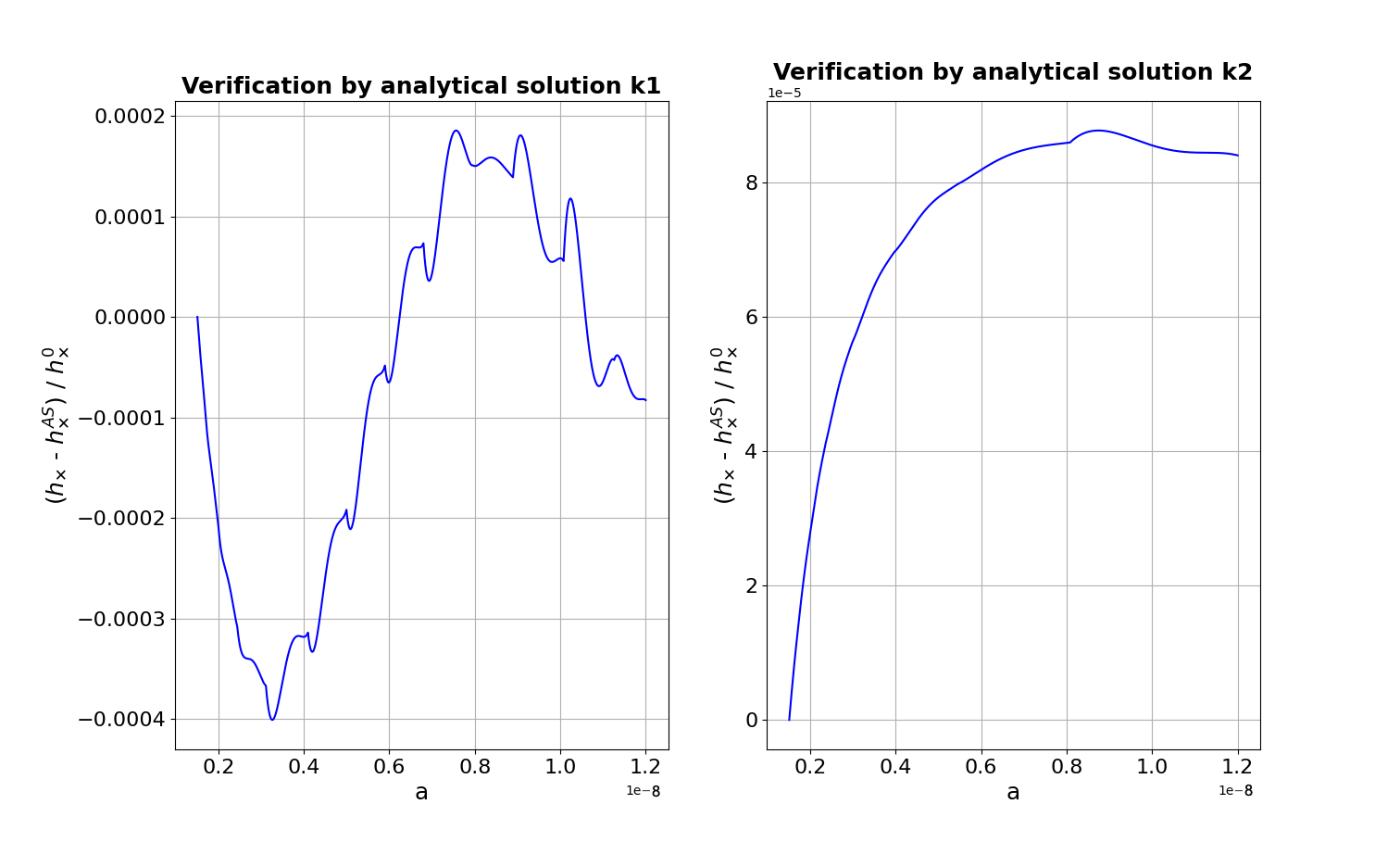}
		\centering
	 	\caption{Absolute difference between the numerical solution and the analytical solution for two frequencies $k_1$ (left panel) and $k_2$ (right panel)}
 		\label{fig:NoPhase}
	\end{figure}

	Here we see a significant discrepancy, but after the correct phase selections, $\phi_0^1=-0.22, \phi_0^2=-0.00065$, we obtain a coincidence with an accuracy of four orders of magnitude (Fig. \ref{fig:sfig2}, \ref{fig:NoPhase}).
	In the Fig.  \ref{fig:NoPhase} an absolute difference between the numerical and the analytical solutions is presented for the two considered cases.
	
	Eventually we can conclude that the method of numerical solution works correctly, and the results obtained for a non-zero value of magnetic field strength are reliable.

	\subsection{Numerical solution results for the system ${\bf\{h_{\times},\,f^x\}}$}
	Let us present the numerical solution results for the system of eqs. (\ref{SDEtoSolve}) with the initial conditions according to the eq. (\ref{InitCond}). The code was written in Python using the solve-ivp package.

{Of particular interest in the long-wave range are the wavelengths that have left their mark on the CMB. We found a solution for frequencies $10^{-18}-10^{-16}$ Hz. 	
We use an implicit Runge-Kutta method of order five to solve the SoDE for  $B_0=1$~ nGs and for these frequencies. 
The scale factor interval is $[10^{-9},\, 10^{-4}]$, and it lies within the radiation dominance (RD) epoch. For comparison, a solution to the system in the absence of a 
magnetic field was also found. }

{The results are the following: by the end of the RD era, amplitude of GW with the selected frequencies is suppressed of about 0.01 percent. Thus, we can conclude that the considered effect of converting GWs into photons in a cosmological magnetic field has an extremely small effect on the amplitude of long-wavelength relict GWs.}

{It is instructive} 
{to say a few words about the physical reason for the  suppression {of the GW amplitude.}
{In the problem we are considering, where the magnetic field is still not strong enough, the main contribution to the damping comes from the classical Maxwell action. Neglecting the loop correction in the equation of motion for the metric perturbation $h_x^y$ in eqs.(\ref{SDEtoSolve}) we obtain the second term in the brackets $\left[\frac{k^2}{a^2}+\frac{16\pi G B_0^2}{a^4}\right]h_{\times}$, which works similarly to the plasma frequency for photons propagating in the plasma \cite{Postnov}. }This term suppresses the low frequency end of the GW spectrum. Indeed, for the quantities $B_0=1$\,nGs, $a_1=10^{-9}$ we obtain
{for the boundary value of the momentum less than the above mentioned analogue of the plasma frequency for GW:}
\be
k \lesssim \frac{\sqrt{16\pi} B_0}{m_{pl}a} \approx \frac{\sqrt{16\pi}\,10^{-9}*1.95*10^{-14}\text{MeV}^2}{2.43*10^{21} \text{MeV} * 10^{-9}}\approx \frac{5.7*10^{-29}\text{eV}}{6.6*10^{-16}\text{eV*s}}\approx 10^{-13}\, \text{Hz}.
\ee}
			

\section{Discussion}
	{In the presented work we {have}
	derived a coupled system {of equations for}
	gravitational and electromagnetic wave propagation in external magnetic field. After that the simplification of the differential equation system was done for the FLRW background metric and for the case of homogeneous magnetic field directed perpendicularly to the initial gravitational wave vector. Finally we {have solved the system numerically}
for $h_{\times}$ - polarization putting  $B_0=1$ nGs}. {The resulting estimate of the effect, without taking into account the inhomogeneity of the magnetic field, is about 0.01\% suppression of the amplitude for a relic GW with a frequency of $10^{-18}$ Hz  at the recombination.}

	{It is worth to note that the results are obtained under 
	{ a large list of simplifying} assumptions and the research 
	demands {deeper} investigation (for example the assumption about the magnetic field homogeneity is  
	{rather crude}). Despite that the results {make sence and one can conclude}
		that the considered phenomenon of GW conversion into photons in the intergalactic magnetic field {cannot} significantly suppress relic gravitational wave amplitude.}
		
		{Let us emphasize that this result  {was not  obvious at the beginning of the research. 
		The smallness of the second-order corrections to the Maxwell action do} not yet mean the smallness of the relic GW suppression effect. It is also necessary to take into account the interaction of the generated by GW photons with the primordial plasma, as well as the fact that the conversion occurs over a long period of time during the evolution of the Universe. A crucial point is also the dependence of the cosmological magnetic field amplitude on the scale factor according to the law $B={B_0}/{a^2}$, which in the early stages of the evolution of the Universe could lead to a rather high magnetic field strength, and therefore to a noticeable conversion effect.}
			
	 {In the future works we plan to solve the second independent part of the SoDE, paying a special attention to the following question: do the emerged scalar perturbations run? After that we want to expand the solution interval up to the end of the matter dominance epoch and to take into account the magnetic field inhomogeneity. }

	{It is worth {stressing} that the stochastic nature of the relic GW direction and the 
	magnetic field direction should have {a large impact on the magnitude}
		of the suppression effect, and {a more} accurate analysis of this {phenomenon} 
		is also very important.  We plan to {perform such}
		analysis in order to {present quantitatively} the dependence of the full relic GW spectrum suppression on the 
		intergalactic magnetic field {strength}.}
		
	{Future research is not only of academic interest, but can also be applied to similar problems of converting gravitational waves into photons near astrophysical sources of strong magnetic fields. Of course, the background metric must be modified to suit the specific task conditions, but the inference structure and some of the qualitative findings discussed in this manuscript will remain valid and useful. In addition, the accuracy of future measurements of CMB polarization \cite{Planck} will steadily increase, and may reach values 
	{of} the order of the considered effect.}

\section*{Author Contributions:}
Conceptualization, A. D. Dolgov; methodology, A. D. Dolgov, L. A. Panasenko; software, L. A. Panasenko.; validation, A. D. Dolgov, L. A. Panasenko and V. A. Bochko; formal analysis, L. A. Panasenko and V. A. Bochko; investigation, A. D. Dolgov, L. A. Panasenko; writing—original draft preparation, L. A. Panasenko; writing—review and editing, A.D. Dolgov, V. A. Bochko; visualization, L. A. Panasenko; supervision, A. D. Dolgov; project administration, A. D. Dolgov; funding acquisition, A. D. Dolgov. All authors have read and agreed to the published version of the manuscript.

\section*{Funding:} 
	The work was supported by the RSF grant number 23-42-00066.
	
\section*{Conflicts of Interest:}
{The authors declare no conflict of interest.}
The funders had no role in the writing of the manuscript; or in the decision to publish the results.

\section*{Abbreviations}
The following abbreviations are used in this manuscript: \\

\noindent 
\begin{tabular}{@{}ll}
GW & Gravitational Wave\\
FLRW space-time & Friedman-LeMaitre-Robertson-Walker space-time\\
EMT & Energy-Momentum Tensor\\
EMW & Electromagnetic Wave\\
HE Lagrangian & Heisenberg-Euler Lagrangian\\
SoDE & System of Differential Equations\\
EoM & Equation of Motion
\end{tabular}


\begin{thebibliography}{99}
\bibitem{g-to-gam1} 
		Gertsenshtein, M.E. Wave Resonance of Light and Gravitational Waves, {\it Zh. Eksp. Teor. Fiz.} {\bf 1961}, 41, 113 [Sov. Phys. JETP, {\bf 1960}, 14, 84].
		
		\bibitem{g-to-gam2}
		Mitskevich, N.V. Fizicheskie polya v obschej teorii otnositel'nosti (Physical fields in General Relativity), { Nauka}: Moscow, 1970.
		
		\bibitem{g-to-gam3}
		Boccaletti, D.; De Sabbata,V. ; Fortini, P.; Gualdi, C. Space-Time Curvature Mode Quanta, {\it Nuovo Cimento}, {\bf 1970}, 70B, 129-146.
		
		\bibitem{g-to-gam4}
		Dubrovich,V.K.  Izvestiya Spetsial'noj Astrofisicheskoj Observatorii, {\bf 1972}, No. 6, 27.
		
		\bibitem{g-to-gam5}
		Zel'dovich,Ya.B. Electromagnetic and gravitational waves in a stationary magnetic field, {\it ZhETF}, {\bf 1973}, 65, 1311 [{\it Sov. Phys. JETP}, {\bf 1974}, 38, 652].
		
		\bibitem{g-to-gam5}
		Fargion, D. Gravitation and Cosmology, {\bf 1995}, 1, 301-310.
		
		\bibitem{g-to-gam6}
		Raffelt, G.; Stodolsky, L.
		Mixing of the Photon with Low Mass Particles,
		{\it Phys.\ Rev.\ D} {\bf 1988}, 37, 1237.
		
		\bibitem{g-to-gam7}
		Dolgov, A.D.; Ejlli, D.  
		Conversion of relic gravitational waves into photons in cosmological magnetic fields, 
		{\it JCAP} {\bf 2012}, 12, 003. 
		
		
		
		
		\bibitem{g-to-gam8}
		Dolgov, A.D.; Ejlli, D.  
		Resonant high energy graviton to photon conversion at post recombination epoch
		{\it Phys. Rev. D.}, {\bf 2013}, 87.104007.
		
		
		
		\bibitem{mtw}
Misner, C.W.; Thorne,  K.S.;  Wheeler, J.A. Gravitation,  W. H. Freeman and company:
San Francisco, 1973.
		\bibitem{GR-2}
		Gorbunov, D.S.; Rubakov, V.A. Introduction to the theory of the early universe: Cosmological perturbations and inflationary theory, Hackensack, USA: World Scientific, 2011. 	
		
		\bibitem{OurWork}
Arbuzova, E.V.; Dolgov,  A.D.; Panasenko, L. A. On graviton propagation in curved spacetime background, {\it J.Exp.Theor.Phys.} {\bf 2022}, 135, 3, 304-311.



		\bibitem{HE}
		Heisenberg, W.; Euler, H. (1936). Folgerungen aus der Diracschen Theorie des Positrons. {\it Zeitschrift für Physik}, {\bf 1936}. 
		Springer Science and Business Media LLC. 98 (11–12): 714–732.
		
		\bibitem{Gasperini}
Fanizza, G.; et al., Linearized propagation equations for metric uctuations in a general (non-vacuum) background geometry, {\it JCAP}, {\bf 2021}, 07.	
		
		\bibitem{LL-2}
Landau, L.D. ; Lifshitz, E.M.  The Classical Theory of Fields, {Volume 2}, Pergamon press: New York, 1971.	

		

		\bibitem{conf-anom}
		Duff, M. J.  Twenty Years of the Weyl Anomaly, {\it Class. Quant. Grav.}{\bf 1994}, 11, 1387 [arXiv:hep-th/9308075]


		\bibitem{Bunch-Davies}
		Bunch, T. S.;  Davies, P. C. W. Stress Tensor and Conformal Anomalies for Massless 		Fields in a Robertson-Walker Universe, {\it Proc. Roy. Soc. London}, {\bf 1978}, A360, 117.
		
		\bibitem{Birrell-Davies}
		Birrell, D.N.; Davies, P. C. W. 
		Quantum Fields in Curved Space,
		Cambridge University Press, 1982. 
	
		\bibitem{AD-anom}
		Dolgov, A. D. 
		Conformal Anomaly and the Production of Massless Particles by a Conformally Flat 			Metric
		{\it Sov. Phys. JETP}, {\bf 1981}, 54, 223 [{\it Zh. Eksp. Teor. Fiz.}, {\bf 1981}, 81,  417].
		
		
		
		\bibitem{AD-anom-2}
		Dolgov, A.D.
		Breaking of conformal invariance and electromagnetic field generation in the universe
		{\it Phys.Rev. D}, {\bf 1993}, 48, 2499-2501 • e-Print: hep-ph/9301280 [hep-ph].

		\bibitem{Parker} 
		Parker, L. {\it Phys. Rev. Lett.}, {\bf 1968}, 21, 562.
		
		\bibitem{Pitaevskii}
		Pitaevskii, L. P.;  Lifshitz, E. M. Physical Kinetic, vol. 10, ch. III
		
		\bibitem{RelPlasma}
		Kraemmer, U.; Rebhan, A. K.; Schulz, H. Resummations in Hot Scalar Electrodynamics, {\it Annals Phys.}, {\bf 1995}, 238, 286
[arXiv:hep-ph/9403301].

		
		
		\bibitem{Lepidi}
		Dolgov,  A.D.;  Lepidi, A.;  Piccinelli, G.
		Electrodynamics at non-zero temperature, chemical potential, and Bose condensate,
		{\it JCAP}, {\bf 2008}, 0902:027,2009; arXiv:0811.4406 [hep-th].
		
		\bibitem{Runge}
		Burrage, K.; Butcher, J. C. Stability Criteria for Implicit Runge-Kutta Methods,		{\it SIAM Journal on Numerical Analysis}, {\bf 1979}, vol. 16, no. 1, pp. 46–57. 
		
				
		\bibitem{MagFieldLimitations}
		Barrow, J. D.;  Ferreira, P. G.;  Silk, J. Constraints on a Primordial Magnetic Field, {\it Phys. Rev. Lett.}, {\bf 1997}, 78, 3610.
		
		\bibitem{Postnov}
Dolgov, A.D.; Postnov, K. Electromagnetic radiation
accompanying gravitational waves
from black hole binaries, {\it JCAP}, {\bf 2017}, 9. 

		
		
		\bibitem{Planck}
		Planck Collaboration, Planck 2018 results. VI. Cosmological parameters, {\it Astron. 			Astrophys.}, {\bf 2020} 641, A6 [arXiv:1807.06209].
		
		\bibitem{Linde}
		Linde, A.D.  Inflationary cosmology, {\it Phys. Repts.}, {\bf 2000}, 333, 17.

		
		
		
		
\bibitem{maggiore-GW}
Maggiore, M.  Gravitational Waves: Volume 1: Theory and Experiments, Oxford University Press, 2008;\\
Gravitational Waves: Volume 2: Astrophysics and Cosmology, Oxford University Press, 2018.
		

\bibitem{Mukha} 
Mukhanov, V.  Physical Foundations of Cosmology, Cambridge University Press: New York, 2005.



\bibitem{SW}
Weinberg, S.  Cosmology, Oxford University Press: 2008.
\bibitem{GWinberg}
Weinberg, S.  Gravitation and Cosmology: Principles and Applns of the General Theory of Relativity, Wiley, 1972.

\bibitem{lifshitz-pert}
Lifshitz, E.M.  On the gravitational stability of the expanding universe, {\it Zh. Eksp. Teor. Phys.}, {\bf 1946}, 15, 587. 

\bibitem{lif-khal}
Lifshitz, E.M.; Khalatnikov, I.M.  Problems of relativistic cosmology, {\it Uspekhi Fizicheskikh Nauk}
 {\bf 1963}, 80, 391.

				
\bibitem{Grishchuk} 
Grishchuk, L.P.  Amplification of gravitational waves in an isotropic universe,{\it Zh. Eksp. Teor. Fiz.} {\bf 1974}, 67, 825; {\it Sov.Phys.JETP}, {\bf 1975}, 40,  409.



\end{thebibliography}
\end{document}